\documentclass[%
 reprint,footinbib,
superscriptaddress,verbatim
 floatfix,
 amsmath,amssymb,
 aps,
prapp, 
longbibliography,
]{revtex4-2}

\usepackage{graphicx}  
\usepackage{dcolumn}   
\usepackage{bm}        
\usepackage{amssymb}   
\usepackage{amsmath}
\usepackage{xcolor}
\usepackage{color}
\usepackage{bbold}
\usepackage{verbatim}
\usepackage[colorlinks=true, pdfstartview=FitV, linkcolor=blue, citecolor=blue, urlcolor=blue]{hyperref} 
\usepackage{tabularx} 
\usepackage{enumitem}
\usepackage{xspace}
\usepackage{braket}
\usepackage{color}
\usepackage{setspace}
\usepackage{mdframed}
\usepackage{framed}
\usepackage{multirow}
\usepackage{setspace}
\usepackage[export]{adjustbox}
\usepackage{varwidth}
\usepackage{float}
\usepackage{verbatim}

\usepackage{lipsum}
\usepackage{physics}
\usepackage[version=3]{mhchem}
\usepackage{color, colortbl}

\makeatletter 
    
\renewcommand\onecolumngrid{
\do@columngrid{one}{\@ne}%
\def\set@footnotewidth{\onecolumngrid}
\def\footnoterule{\kern-6pt\hrule width 1.5in\kern6pt}%
}

\renewcommand\twocolumngrid{
        \def\footnoterule{
        \dimen@\skip\footins\divide\dimen@\thr@@
        \kern-\dimen@\hrule width.5in\kern\dimen@}
        \do@columngrid{mlt}{\tw@}
}%

\makeatother    

\newcommand{\nocontentsline}[3]{}

\renewcommand{\i}{\mathrm{i}}

\def\equationautorefname~#1\null{Eq.~(#1)\null}

\newcommand{\iu}{\mathrm{i}}
\newcommand{\om}{\omega_\text{m}}
\newcommand{\phim}{\phi_\text{m}}

\def\equationautorefname~#1\null{Eq.~(#1)\null}

\begin{document}

\title{Optomechanical realization of the bosonic Kitaev-Majorana chain}

\author{Jesse J.~Slim}
\affiliation{Center for Nanophotonics, AMOLF, Science Park 104, 1098 XG Amsterdam, The Netherlands}
\affiliation{ARC Centre of Excellence for Engineered Quantum Systems, School of Mathematics and Physics, University of Queensland, St. Lucia, QLD 4072, Australia.}

\author{Clara C.~Wanjura}
\affiliation{Max Planck Institute for the Science of Light, Staudtstraße 2, 91058 Erlangen, Germany}

\author{Matteo Brunelli}
\affiliation{Department of Physics, University of Basel, Klingelbergstrasse 82, 4056 Basel, Switzerland}

\author{Javier del Pino}
\affiliation{Institute for Theoretical Physics, ETH Zürich, 8093 Zürich, Switzerland}
\affiliation{Center for Nanophotonics, AMOLF, Science Park 104, 1098 XG Amsterdam, The Netherlands}

\author{Andreas Nunnenkamp}
\affiliation{Faculty of Physics, University of Vienna, Boltzmanngasse 5, 1090 Vienna, Austria}

\author{Ewold Verhagen}
\email{verhagen@amolf.nl}
\affiliation{Center for Nanophotonics, AMOLF, Science Park 104, 1098 XG Amsterdam, The Netherlands}

\email[Corresponding author: ]{verhagen@amolf.nl}

\begin{abstract}
The fermionic Kitaev chain is a canonical model featuring topological Majorana zero modes. 
We report the experimental realization of its bosonic analogue in a nano-optomechanical network where parametric interactions induce two-mode squeezing and beamsplitter coupling among the nanomechanical modes, equivalent to hopping and superconductor pairing in the fermionic case, respectively.
We observe several extraordinary phenomena in the bosonic dynamics and transport, including quadrature-dependent chiral amplification, exponential scaling of the gain with system size, and strong sensitivity to boundary conditions.
Controlling the interaction phases and amplitudes uncovers a rich dynamical phase diagram that links the observed phenomena to non-Hermitian topological phase transitions.
Finally, we present an experimental demonstration of an exponentially enhanced response to a small perturbation as a consequence of non-Hermitian topology.
These results represent the demonstration of a novel synthetic phase of matter whose bosonic dynamics do not have fermionic parallels, and establish a powerful system to study non-Hermitian topology and its applications in signal manipulation and sensing.
\end{abstract}

\maketitle

Topological phases of matter have revolutionized our understanding of electronic materials \cite{Hasan2010} and classical wave systems \cite{Ozawa2019,Ma2019}, with unique applications in robust information transmission, metrology, and quantum computing~\cite{Nayak2008,Beenakker2013}.
Intense interest focused recently on topological phenomena in metamaterials that feature non-Hermitian dynamics --- associated with amplification and dissipation that are ubiquitous in bosonic domains such as photonics and acoustics \cite{Bergholtz2021,Ding2022nonhermitian}.
On the one hand, there is the fundamental question of how to recognize and classify non-Hermitian topological phases.
Indeed, whereas topological invariants in Hermitian systems are usually defined on the structure of their eigenvectors~\cite{Hasan2010}, non-Hermitian systems can additionally be classified on the structure of their complex eigenvalues \cite{Gong2018,Kawabata2019topological,Kawabata2019symmetry}. 
On the other hand, the merger of non-Hermitian dynamics and topology may lead to new physical phenomena, and associated applications from robust lasing and topologically protected amplification to enhanced sensing~\cite{Peano2016topologicalQuantum,Bandres2018,Budich2020}.
A widely studied phenomenon is the non-Hermitian skin effect (NHSE); the accumulation of a macroscopic number of states at boundaries of non-Hermitian lattices \cite{Yao2018edge,Yao2018nonhermitian,Kunst2018biorthogonala,MartinezAlvarez2018nonhermitian,Okuma2020topological,Zhang2020correspondence}. 
This was observed experimentally in various systems~ \cite{Ghatak2020observation,Helbig2020generalized,Weidemann2020topological,Xiao2020nonhermitian,Liang2022dynamic,Zhang2021acoustic,Wang2021generating,Wang2022nonhermitian} and seemingly violates the bulk-boundary correspondence that famously links the existence of robust edge states to the topological description of a material's bulk properties in Hermitian topological insulators.

\begin{figure*}[t]
    \includegraphics[width=\textwidth]{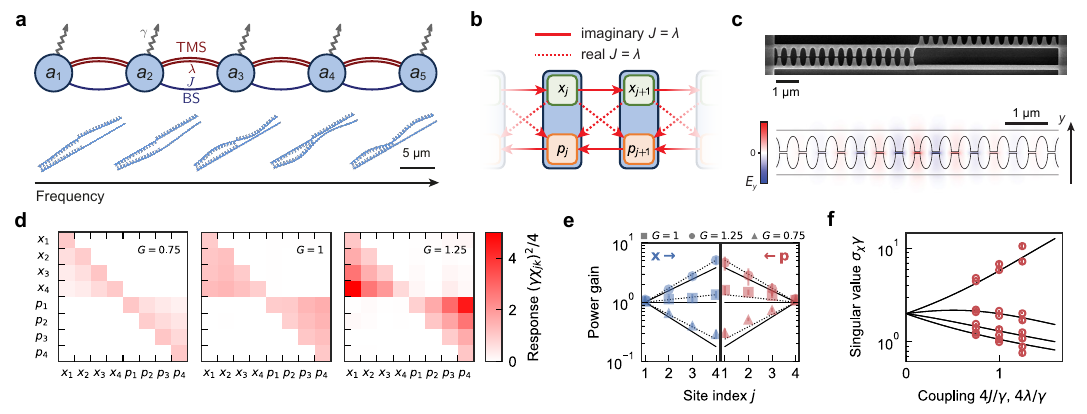}
    \caption{\textbf{Bosonic Kitaev-Majorana Chain (BKC) in an optomechanical network.}
    (\textbf{a}) The BKC comprises resonators coupled through beamsplitter (BS) and two-mode squeezing (TMS) interactions. The mechanical displacement profiles of the modes that form the sites of the chain are depicted below.
    (\textbf{b}) By tuning the phases of BS and TMS amplitudes, specific quadratures can be coupled in a unidirectional fashion for $|J|=|\lambda|$.
    (\textbf{c}) SEM micrograph of the sliced photonic crystal nanobeam system (top) and simulated electric field in the nanocavity that is coupled to the mechanical modes (bottom). The time-modulated cavity field induces all parametric interactions.
    (\textbf{d}) Susceptibility matrix $\chi_{jk}$ resolved in the resonator quadratures along the chain for purely imaginary $J=\lambda=\iu \mu$. As the per-link gain $G=4\mu/\gamma$ increases, the system features quadrature-dependent chiral end-to-end amplification.
    (\textbf{e}) Concomitantly, the chain builds up exponential gain along the decoupled chains of $x_j$ and $p_j$ quadratures. The expected gain profiles for the nominal values of $G$ noted above are shown (solid lines). A closer agreement is obtained after including a correction of $+5\%$ on all interaction rates (dashed lines).
    (\textbf{f}) Singular values $\sigma_\chi$ of the susceptibility matrices in (\textbf{d}) as a function of $G$, corresponding to gain of the various transmission channels. The emerging separation between two (amplifying) singular values and the remaining ones with much smaller gain can be associated with a non-Hermitian bulk-boundary correspondence. Error bars represent $\pm2\sigma$ with $\sigma$ the standard deviation obtained by repeating each experiment $10$ times.}
    \label{fig:1}
\end{figure*}

Against this backdrop, McDonald et al.~recently proposed the bosonic Kitaev-Majorana chain (BKC) \cite{McDonald2018}. It is a bosonic analogue of the fermionic Kitaev chain --- the well-known one-dimensional model that predicts topologically-protected Majorana zero modes at the ends of a superconducting wire \cite{Kitaev2001unpaired}.
The BKC is formed by coupling bosonic modes through both beamsplitter and two-mode squeezing interactions, mimicking hopping and $p$-wave superconducting pairing in the fermionic Kitaev chain, respectively (Fig.~\ref{fig:1}).
While the BKC lacks the fermionic statistics that make Majorana bound states interesting to quantum computing~\cite{Nayak2008,Beenakker2013}, it hosts a set of remarkable features, that also set it apart from other metamaterial networks~\cite{Chen2019,Qian2023}. 
These features include quadrature-dependent chiral transport and amplification, and strong sensitivity to boundary conditions which can be connected to its non-Hermitian topological properties~\cite{McDonald2018,Wanjura2020}.
The BKC has been predicted to offer sensing applications due to an exponential boost of sensitivity to signals perturbing the end of the chain \cite{McDonald2020}, and metastable states at the chain ends have been described as bosonic Majorana zero modes \cite{Flynn2021}. All of these traits establish the BKC as a highly interesting non-Hermitian topological phase of matter.

\subsection*{Optomechanical resonator chain}

We realize the bosonic Kitaev-Majorana chain experimentally in an actively controlled classical optomechanical network, study the rich dynamical behavior that ensues, and link it to the system's non-Hermitian topology. 
The nodes of the one-dimensional chain are the flexural nanomechanical modes of a pair of silicon strings (\autoref{fig:1}a,c) with different resonance frequencies $\omega_j/(2\pi)$ ranging between 3.7 and 26~MHz. The displacements of all mechanical modes affect the optical frequency of a single photonic nanocavity defined in the nanoscale gap between the strings by the sliced photonic crystal geometry~\cite{mathew2018synthetic,delpino2021nonhermitian}. This optomechanical coupling implies that radiation pressure of a drive laser that is coupled to the cavity from free space and tuned to the side of the cavity resonance (linewidth $\kappa/2\pi=320$~GHz) induces a strong optomechanical spring effect, as the optical force adiabatically follows mechanical displacement (in the `bad cavity' regime, $\kappa\gg\omega_j$). Temporal modulation of the laser intensity then yields control over the nanomechanical interaction Hamiltonian through parametric driving: Beamsplitter (BS) interactions between modes $i$ and $j$ are induced by modulation at the respective frequency difference $|\omega_j-\omega_i|$, and two-mode squeezing (TMS) through sum-frequency modulation at $|\omega_i+\omega_j|$, with each coupling's rate and phase controlled by the depth and phase of the relevant modulation tone~\cite{supplementary,delpino2021nonhermitian,Wanjura2023}.
The effective Hamiltonian that we implement reads
\begin{align}
    H_\text{BKC} = \sum_j \left( J a_{j+1}^\dagger a_j + \lambda a_{j+1}^\dagger a_j^\dagger + \text{H.c.} \right), \label{bkc:eq:general_hamiltonian}
\end{align}
with modes described by bosonic annihilation operators $a_j$ in frames rotating at $\omega_j$, and $J$ ($\lambda$) the complex BS (TMS) amplitude, set identical on each link of the chain. While this Hamiltonian is obviously Hermitian, it generates non-Hermitian dynamics in the modes $a_j$, associated with the particle-non-conserving TMS interactions.

\begin{figure*}[t]
    \includegraphics[width=12cm]{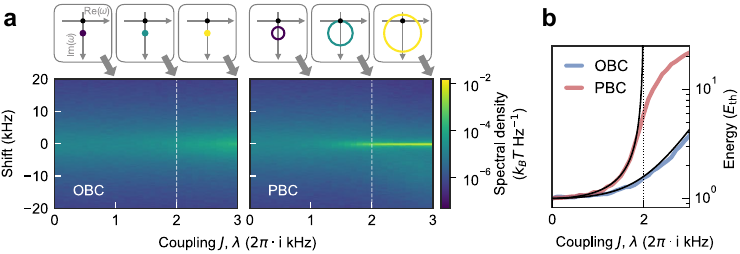}
    \caption{\textbf{Boundary-dependent instability.}
    (\textbf{a}) Thermomechanical spectra for $J = \lambda = i \mu$ as a function of frequency and coupling strength for open boundary conditions (OBC) and periodic boundary conditions (PBC).
    The row above sketches the complex eigenvalues of the dynamical matrix under PBC and OBC, respectively. Under PBC there are two degenerate complex bands featuring a point gap with opposite winding. As $\mu$ increases, the circle radius becomes bigger, and eventually the band winds around the origin (marked by dashed line). The system under PBC then becomes dynamically unstable. The situation is very different under OBC where the spectrum collapses to an exceptional point and no instability is observed in the linear response.
    (\textbf{b}) Total thermal energy in the first resonator as a function of coupling strength. Beyond the instability observed for PBC, nonlinearities limit the self-oscillation amplitude~\cite{supplementary}. Under OBC the linear theory continues to describe the system well.
    } \label{fig:2}
\end{figure*}

\subsection*{Chiral amplification}
We measure the chain's full response to a modulated radiation pressure drive resonant with any of the modes, by demodulating the induced vibrations detected by a second (readout) laser.
Similar to its fermionic counterpart, the BKC is best understood using Hermitian quadrature operators $x_j=( a_j + a_j^\dagger )/\sqrt{2}$ and $p_j= ( a_j - a_j^\dagger ) / (\iu \sqrt{2})$.
We separately resolve the $x$ and $p$ quadratures~\cite{Wanjura2023,supplementary} in the measured susceptibility matrix $\chi$ shown in Fig.~\ref{fig:1}d, which relates the steady-state amplitudes $\mathbf{q} = (\langle x_1\rangle,\dots, \langle x_N\rangle,\langle p_1\rangle,\dots,\langle p_N\rangle)^T$ (in units of zero-point amplitudes $x_{\text{zpf},j}$) to resonant drives of each quadrature $\mathbf{f}^{(\mathbf{q})} = (f_{x_1},\dots, f_{x_N}, f_{p_1}, \dots, f_{p_N})$ through $\mathbf{q} = \chi \mathbf{f}^{(\mathbf{q})}$~\cite{supplementary}.
We choose in Fig.~\ref{fig:1}d both $J$ and $\lambda$ to be purely imaginary and equal; $\lambda=J=\iu \mu$ ($\mu>0$), and plot $\chi^2$ for $N=4$ for various values of $G\equiv4\mu/\gamma$. The damping rates of all modes are equalized to $\gamma_j/(2\pi) = \gamma/(2\pi) = 8$~kHz through measurement-based feedback~\cite{supplementary}. Drives are normalized such that $x=2f_x/\gamma$ for a single, uncoupled resonator. 
We observe different response for different quadratures, related to interference of the BS and TMS processes~\cite{Metelmann2015,McDonald2018,Wanjura2023}. In fact, the chosen phase of $J$ fully decouples $x_j$ and $p_j$ quadratures (Fig.~\ref{fig:1}b)~\cite{McDonald2018,supplementary}, with $x$ signals propagating rightward and $p$ leftward. Such a dual character of chiral (unidirectional) transport invites comparison to the two edge states in the quantum spin Hall effect, whose directions are coupled to opposite spins~\cite{Hasan2010}.
We recognize that for the balanced condition of equal hopping and squeezing strengths $|\lambda|=|J|$, the phase-dependent transport is fully unidirectional.
Notably, the quadratures $x_N$ and $p_1$ are isolated from the rest of the chain, reminiscent of the fermionic Kitaev chain: There, balancing hopping and pairing completely decouples the unpaired Majorana zero modes at the ends of the chain. 

Importantly, for coupling amplitudes where $G>1$ we observe signal amplification, with exponential growth in opposite directions for orthogonal quadratures (Fig.~\ref{fig:1}e). This confirms the prediction that the BKC serves as a unique type of \emph{directional amplifier}~\cite{McDonald2018}.
The susceptibility matrix derived from the Hamiltonian~(\ref{bkc:eq:general_hamiltonian}) including dissipation~\cite{supplementary} reads for $J=\lambda=\iu \mu$
\begin{align}
|\chi| = 
\left|\begin{matrix}
    \chi_{xx} & \chi_{xp} \\
    \chi_{px} & \chi_{pp}
\end{matrix}\right|
=\frac{2}{\gamma}\begin{pmatrix}
1 \\
G & 1 & & & & &\text{\huge0} \\
G^2 & G & 1 \\
G^3 & G^2 & G & 1 \\
			& & & & 1 & G & G^2 & G^3 \\
			& & & & & 1 & G & G^2 \\
			& \text{\huge0} & & & & & 1 & G \\
			& & & & & &  & 1
\end{pmatrix}.
\end{align}
We recognize the quadrature decoupling in the block-diagonality of this matrix and characteristic exponential scaling of gain with system size, with $G$ the gain per link.
To identify the independent transmission channels, we plot the singular values of the measured susceptibility matrices in Fig.~\ref{fig:1}f. This reveals precisely two amplification channels associated with the largest singular values, which separate from the other components as the gain $G$ (or the system size) increases.

\subsection*{Non-Hermitian topology}
The unique set of transport properties highlighted above are rooted in topology, although the BKC's topological nature markedly differs from the fermionic case. It can be associated with the complex eigenvalues of the chain, which read for purely imaginary $J,\lambda$ under periodic boundary conditions~\cite{supplementary}
\begin{equation}
    \omega_{x,p}(k)=-\mathrm{i}\frac{\gamma}{2}-2J\sin k \mp 2\mathrm{i}\lambda\cos k,\label{eq:spectrum}
\end{equation}
with wavevector $k$ and choosing $+$ ($-$) for $x$ ($p$).
These two eigenvalues are associated with the two blocks $\chi_{xx}$ and $\chi_{pp}$ of the susceptibility matrix and each equal to those of the Hatano-Nelson model; the canonical non-Hermitian model that describes a chain with asymmetric hopping amplitudes
~\cite{Hatano1997vortex}. Indeed, the block-diagonal non-Hermitian BKC Hamiltonian can be interpreted (up to an imaginary shift) as two copies of the Hatano-Nelson model for different quadratures~\cite{McDonald2018,supplementary}.
The spectrum under periodic boundary conditions (\ref{eq:spectrum}) correspondingly features a \emph{point gap}: a region in the complex plane that is enclosed by a non-Hermitian band, to which a \emph{spectral winding number} can be assigned as topological invariant \cite{Gong2018,Ding2022nonhermitian}.

\begin{figure*}[t]
    \includegraphics[width=12cm]{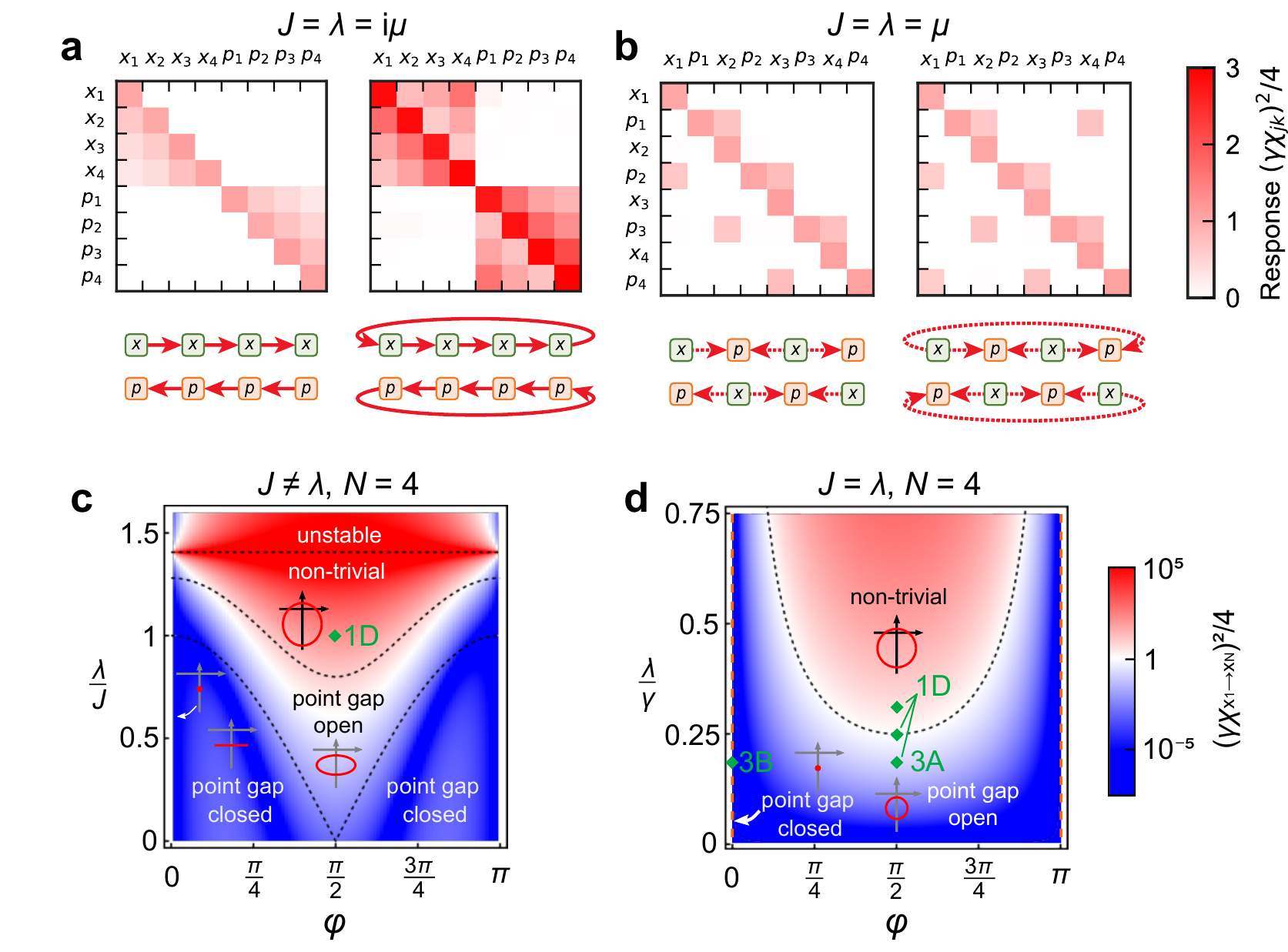}
    \caption{\textbf{Transport properties and topological phase diagrams of generalized bosonic Kitaev-Majorana chains.}
    (\textbf{a},\textbf{b}) Measured susceptibility matrices for $N = 4$. In (\textbf{a}) $\lambda=J=\iu \mu$ for OBC (left) and PBC (right). In (\textbf{b}) $\lambda=J=\mu$ for OBC (left) and PBC (right). In all cases $\mu/(2\pi) = 1.5$ kHz and $\gamma/(2\pi) = 8$ kHz such that $G=0.75$.
      The sketches below the panels indicate the quadrature-dependent couplings that underlie the transport behavior.
    (\textbf{c}) The generalized BKC features four different regimes, distinguished by the spectral structure under PBC: point gap closed, point gap open without spectrally winding the origin, a regime of non-trivial non-Hermitian topology where the spectrum winds the origin, and a fully unstable regime. Dashed lines indicate phase boundaries.
    The color shows end-to-end gain $\chi_{x_1\rightarrow x_N}$ in the corresponding open chain as a function of phase $\varphi$ and $\lambda/J$ for $J/\gamma = 5/16$ and $N = 4$.
    (\textbf{d}) For $\lambda=J$ the phase diagram simplifies: The OBC system is guaranteed to be stable and a point gap opens except at $\varphi=0,\pi$ (red dashed lines). Green diamonds indicate the parameters of the experiments in Figures 1 and 3.
    }\label{fig:3}
\end{figure*}

A point-gapped spectrum is accompanied by the NHSE in the corresponding finite (open boundary) chain, which implies the breakdown of the bulk-boundary correspondence that links the number of edge states to the topological invariant~\cite{Bergholtz2021,MartinezAlvarez2018nonhermitian,Gong2018,Xiong2018why,Wanjura2020,Okuma2020topological,Zhang2020correspondence,Coulais2021,Ding2022nonhermitian,Wang2021generating,Brunelli2022restoration}. 
Bulk-boundary correspondence can be restored by specifically considering spectral windings around the origin of the complex plane~\cite{Porras2019,Wanjura2021Correspondence,Brunelli2022restoration}.
That winding number acts as a topological invariant that predicts the number of singular values of the non-Hermitian Hamiltonian matrix $\chi^{-1}$ that split off from the bulk values and approach zero for large system size~\cite{Brunelli2022restoration}. 
Those zero singular values are directly connected to diverging singular values of the susceptibility $\chi$. The two channels of directional amplification that we recognized in Fig.~\ref{fig:1}f thus have a topological nature, and can be compared to the zero modes of the fermionic Kitaev-Majorana chain.

At variance with the Hatano-Nelson model, the non-Hermitian topological phase of the BKC is protected by a parity symmetry~\cite{Kawabata2019symmetry}, which also decouples the quadratures and doubles the eigenvalues in eq.~\ref{eq:spectrum}.
This symmetry expresses the fact that the system is invariant under reflection and exchanging $x$ and $p$~\cite{supplementary} (Fig.~\ref{fig:1}d).
Interestingly, the fact that our experiment employs a synthetic dimension means that detuning disorder can be minimized through careful adjustment of the modulation frequencies. This ensures the symmetry, revealing the non-Hermitian topology and all its implications in our optomechanical BKC.

A distinctive trait of non-Hermitian topological models is sensitivity to boundary conditions. 
We realize periodic boundary conditions (PBC) in our optomechanical network by connecting also the last node to the first. 
As the gain exceeds unity, we observe strong increase accompanied by spectral narrowing of the thermal fluctuations that is characteristic of dynamical instability for PBC (Fig.~\ref{fig:2}). The oscillation amplitude is limited by nonlinearities~\cite{supplementary}. In stark contrast, the system under open boundary conditions (OBC) is dynamically stable at the same gain.
This boundary-dependent dynamical instability is thus a dramatic feature that coincides for our zero-detuning network with the non-Hermitian topological phase transition that occurs as soon as the complex energies under PBC start to wind the origin~\cite{Porras2019,Brunelli2022restoration}. 
This is a convective instability, i.e.~excitations experience net round-trip gain and grow indefinitely around the closed chain.
For the open chain, the point gap remains closed and the dynamics stable (Fig.~\ref{fig:2}): Amplified excitations reach the end of the chain, but cannot reflect~\cite{McDonald2018}.

\begin{figure*}
    \includegraphics[width=12cm]{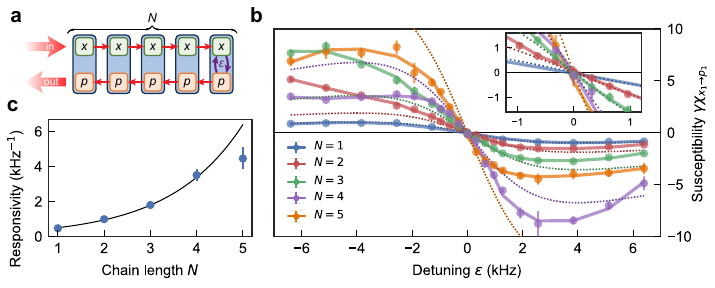}
    \caption{\textbf{Experimental demonstration of exponentially-enhanced sensing.} (\textbf{a})~The response of a BKC comprising $N$ resonators with interactions $J = \lambda = \iu \mu$ is very sensitive to the local detuning $\varepsilon$ of the last site. To sense small $\varepsilon$, a coherent drive of the first site $x_1$ gets amplified towards the right. The perturbation $\varepsilon$ converts a fraction of the excitation to the orthogonal quadrature $p_N$ and gets amplified back towards the first site, where it is detected~\cite{McDonald2020}. The response $\chi_{x_1\to p_1}(\varepsilon)$ is thus a measure for $\varepsilon$, with the chain providing amplification. (\textbf{b})~Susceptibility $\chi_{x_1\to p_1}(\varepsilon)$ of $p_1$ to a signal injected in $x_1$ as a function of $\varepsilon$ in experiment (circles), with the inset zooming on small $\varepsilon$. Here, $\mu/(2\pi) = 2.75$ kHz, $\gamma/(2\pi) = 8$ kHz (per-link gain $G = 1.37$). We attribute the deviation between measured response and linear theory (dashed lines) for large $\varepsilon$ to nonlinear effects~\cite{supplementary}. (\textbf{c})~Linear responsivity $\mathcal{R}$ to small $\varepsilon$ for increasing $N$ extracted form (\textbf{b}). We observe exponential increase with $N \leq 4$, as predicted by the linear theory (line) using the independently estimated $G=1.37$. The decreased responsivity for $N=5$ is attributed to nonlinearity. Error bars represent $\pm2\sigma$ with $\sigma$ the standard deviation from $5$ experiments.} \label{fig:4}
\end{figure*}

\subsection*{Topological phase diagram}
So far we focused on the choice $\varphi=\pi/2$ for the BS and TMS interactions $J=|J|e^{\iu \varphi}$ and $\lambda = |\lambda|e^{\iu \varphi}$, respectively, for which connections to both the fermionic Kitaev-Majorana and non-Hermitian Hatano-Nelson models are most direct.
Given the active phase control of our platform, we explore the behavior for different phases $\varphi$. 
Figures~\ref{fig:3}a,b compare the effect of open or closed boundaries on the response for $\varphi=\pi/2$ and $\varphi=0$, respectively, in the stable regime ($G=0.75$). For $\varphi=\pi/2$ the sensitivity to boundary conditions, associated with the opened point gap and NHSE, is clear again: chiral quadrature-dependent transport is terminated at the ends of a finite chain, whereas for PBC signals can circulate continuously, opening transmission to all sites. In contrast, when $\varphi = 0$ (real $J = \lambda$, Fig.~\ref{fig:3}b), an injected signal only elicits a response in neighboring sites, regardless of boundary conditions. This means that also dynamical stability no longer depends on boundary conditions and the chain is topologically trivial for any $G$.
The behavior can be understood from the fact that $x$ and $p$ quadratures no longer cascade; destructive interference between BS and TMS processes leads to a decoupling of dimers. 

In fact, a rich phase diagram emerges as control phases and strengths are varied. 
As the BKC preserves parity symmetry for all $\varphi$, its eigenvalues under PBC always form two degenerate bands in the complex plane.
We recognize four distinct dynamical phases in Fig.~\ref{fig:3}e: 
The squeezing-induced point gap~\cite{Wan2023} is opened when $|\lambda| > |J| |\cos(\varphi)|$. 
For sufficiently small dissipation, i.e., $\gamma<\sqrt{|\lambda|^2-|J|^2\cos^2\varphi}$, the spectrum winds the origin, marking the topological transition to the non-trivial regime and the boundary-dependent instability that we observed in Fig.~\ref{fig:2}. The open chain then functions as a directional amplifier. 
Chains with open point gap can always be interpreted as two Hatano-Nelson chains of opposite winding, as a transformation maps them for any $\varphi$ to the case $\varphi=\pi/2$ (with reduced interaction strengths) \cite{McDonald2018,supplementary}.
In contrast, the regime  $|\lambda|<|J| |\cos(\varphi)|$ is characterized by a closed point gap with eigenstates that are largely insensitive to boundary conditions, and one can map the system to a trivial hopping chain~\cite{McDonald2018,supplementary}.
Finally, also the open chain becomes unstable when 
$\lvert\lambda/J\rvert>(1+\left(\gamma/(2J)\right)^2\sec^2\left(\pi N/(1+N)\right))^{1/2}$.

The phase diagram simplifies in the special case $|\lambda| = |J| = \mu$ (Fig.~\ref{fig:3}d). 
The topological phase transition to the directional amplification regime then happens when $2\mu\sin \varphi > \gamma/2$, while the closed point gap region collapses onto lines at $\varphi=0,\pi$ that are independent of $\mu$. There, all PBC eigenvalues coalesce to a single value $\epsilon_0 = -\iu\gamma/2$, and the dynamical matrix of a finite closed chain exhibits $N$ degenerate order-2 exceptional points. This special case was probed in Fig.~\ref{fig:3}b, where we saw that excitating the $x$ quadrature in one site only elicits a $p$ response of its direct neighbors, while $p\rightarrow x$ is forbidden due to destructive interference, an interesting form of quadrature nonreciprocity~\cite{Wanjura2023,supplementary}. 
In these examples, we see that the BKC's non-Hermitian bandstructure thus links to both its dynamics and transport. 

\subsection*{Exponentially-enhanced perturbation responsivity}
Unlike the phase $\varphi$, a finite detuning on any of the sites breaks the parity symmetry protecting the BKC's non-Hermitian topology~\cite{supplementary}. 
Indeed, detuning a single site, in particular at an edge, significantly alters the responsivity. This setting has been proposed as a sensor with exponentially-enhanced sensitivity~\cite{McDonald2020}. 

Fig.~\ref{fig:4}a illustrates the working principle of a sensor exploiting directional amplification in the BKC, i.e. 
an $N$-site open chain with imaginary couplings $J$, $\lambda$. 
In a non-trivial chain, a signal injected at the $x_1$ quadrature is amplified to the other end. 
At the last site, a detuning $\varepsilon$ --- the quantity to be sensed --- converts the excitation in $x_N$ to the orthogonal quadrature $p_N$. Subsequently, the converted signal is again amplified towards the $p_1$ quadrature of the first site and detected.
Importantly, quadratures $x_N$ and $p_N$ \emph{only} couple due to the presence of the detuning $\varepsilon$.
Thus, a measurement of $\chi_{x_1\to p_1}$ is highly sensitive to any perturbation that dispersively couples to the last resonator, such as for example the mass of an adhered protein or the state of a qubit coupled to the resonator~\cite{McDonald2020}.

We implement the sensing scheme with  $J=\lambda=\mathrm{i}\mu$ in nanomechanical chains comprising up to $5$ sites. A detuning $\varepsilon$ at the last site is controlled via tuning of the relevant pump tones, serving as a proxy for a dispersively coupled perturbation.
We show the measured response $\gamma\chi_{x_1\to p_1}(\varepsilon)$ to a drive of $x_1$ in Fig.~\ref{fig:4}b, revealing a strong dependence on $\varepsilon$. We note that the slope of $\gamma\chi_{x_1\to p_1}(\varepsilon)$ close to $\varepsilon=0$ gets steeper for larger system size $N$ (Fig.~\ref{fig:4}b, inset).
We plot this experimentally retrieved linear responsivity~\cite{supplementary}
\begin{align}
    \mathcal{R} \equiv \gamma \left\lvert\frac{\partial \chi_{x_1\to p_1}}{\partial\varepsilon}\right\rvert_{\varepsilon=0}.
\end{align}
as a function of system size in Fig.~\ref{fig:4}c. Notably, the responsivity grows \emph{exponentially} with the chain length $N$. 
We can understand this based on the non-Hermitian topological, directional amplifier (Fig.~\ref{fig:1}), since in topologically non-trivial phases it displays a susceptibility $\lvert\chi_{x_1\to x_N}(\varepsilon=0)\rvert = \lvert\chi_{p_N\to p_1}(\varepsilon=0)\rvert\propto e^{\alpha N}$ with some $\alpha>1$, that scales exponentially with system size~\cite{Wanjura2020,McDonald2020}. Concretely~\cite{supplementary}, the back-and-forth amplification mechanism implies $\mathcal{R} =\gamma [\chi_{x_1 \to x_N}(\varepsilon=0)]^2$.
Crucially, a point gap is not sufficient for exponential enhancement; it \emph{only} occurs for non-trivial winding number as $\chi_{x_1\to x_N}$ is exponentially attenuated otherwise~\cite{supplementary}.
For equal coupling strengths, the responsivity takes on the simple form $\mathcal{R} = 4G^{2(N-1)}/\gamma$.

For large $\varepsilon$ or large total gain, the increased amplitude in the chain prompts non-linear effects that introduce additional detunings and saturate the per-link gain, due to a relatively strong optomechanical nonlinearity~\cite{supplementary}. This causes deviation from linear theory in Fig.~\ref{fig:4}b and a slightly smaller than predicted enhanced sensitivity for $N=5$ in Fig.~\ref{fig:4}c.
The effect of potential (non-linear) detuning on other sites than the sensing site will be interesting to study in depth in light of practical sensor implementations. Likewise, while effective enhancement of quantum Fisher information through the demonstrated mechanism has been predicted~\cite{McDonald2020}, careful evaluation of the contributions of thermal or quantum fluctuations should be the subject of future work.

\subsection*{Conclusions and outlook}
These results point to potential applications of the BKC in sensing and directional amplification, which are both related to its behavior as a unique non-Hermitian topological phase. Many more directions of study are opportune: characterizing the properties of the chain's Majorana zero modes~\cite{Flynn2021}, exploring the inclusion of nonlinearity and associated `fermionization' of the system~\cite{Bardyn2012}, extension to higher dimensions, and studying non-Hermitian braiding operations~\cite{Barlas2020,Patil2022}. We have experimentally shown that the BKC constitutes a non-Hermitian topological model that unifies the concepts of spectral winding, the non-Hermitian skin effect, and directional amplification, with different topological phases linked to observable effects in stability and transport. Importantly, implementations of the BKC are not limited to optomechanics; realizations can be envisioned in electronics, acoustics, photonics, and superconducting circuits. In a fully quantum setting, it would be interesting to study the behavior of fluctuations and the generation of many-body entanglement~\cite{Lee2023} --- relevant to the use of the BKC and related phases in quantum sensing and information processing. 

\subsection*{Acknowledgements}
We acknowledge A.~A.~Clerk for discussions. M.B.~acknowledges funding from the Swiss National Science Foundation (PCEFP2\_194268). J.d.P.~acknowledges financial support from the ETH Fellowship programme (grant no.~20-2 FEL-66). This work is part of the research programme of the Netherlands Organisation for Scientific Research (NWO). It is supported by the European Research Council (ERC starting grant no.~759644-TOPP).


%


\clearpage
\pagebreak
\renewcommand{\theequation}{S\arabic{equation}}
\renewcommand{\thefigure}{S\arabic{figure}}
\setcounter{equation}{0}
\setcounter{table}{0}
\renewcommand{\figurename}{FIG.}
\setcounter{figure}{0}
\renewcommand{\theHfigure}{S.\thefigure}

\patchcmd{\section}{\centering}{\raggedright}{}{}
\patchcmd{\subsection}{\centering}{\raggedright}{}{}
\patchcmd{\subsubsection}{\centering}{\raggedright}{}{}


\newpage
\onecolumngrid
\part*{\large\centering Supplementary Information}

\section{Supplementary experimental details: Controlling the BKC interaction Hamiltonian and measurement of quadrature-resolved susceptibility}

We use the techniques outlined in~\cite{mathew2018synthetic_SI,delpino2021nonhermitian_SI,Wanjura2023_SI}
to synthesize arbitrary quadratic bosonic Hamiltonians in a network of nanomechanical modes controlled through temporally modulated radiation pressure of control lasers. In the following, we briefly outline the main working principles and methods that we use to implement the Hamiltonian and read out the response of the system.

\subsection{Experimental system}
The nano-optomechanical device used in this study comprises a suspended silicon sliced nanobeam featuring a photonic crystal cavity (optical mode frequency $\omega_\text{c}/(2\pi) = 195.451$ THz, linewidth $\kappa/(2\pi) = 320$ GHz) coupled to multiple non-degenerate, flexural mechanical modes. Details on design, fabrication, and characterization of the device can be found in~\cite{delpino2021nonhermitian_SI,Wanjura2023_SI}.
We use the lowest five mechanical modes (`resonators') of the set of strings that compose the nanobeam, with spatial profiles as shown in Fig.~1a of the main text and frequencies $\omega_j/(2\pi) = \{3.7, 5.3, 12.8, 17.6, 26.3\}$ MHz, to construct the chains studied in this work. These mechanical resonators have intrinsic linewidths $\gamma_j/(2\pi) \approx 1$--$7$ kHz and couple dispersively to the optical cavity mode with vacuum optomechanical coupling $g_{0,j}/(2\pi) \approx \{ 5.3, 5.9, 3.3, 3.1, 1.9 \}$ MHz. These are defined as the cavity frequency shift induced by a mechanical displacement of the respective mode equal to the magnitude of its ground state fluctuations~\cite{Aspelmeyer2014_SI}. 
Light is coupled in and out of the cavity from free space, by focusing a laser beam at normal incidence to the sample surface, in a cross-polarized detection scheme. The coupling efficiency is approximately 3\%, and typical incident laser powers are of the order of milliwatts. The sample is positioned in a vacuum chamber (pressure $2\times10^{-6}$~mbar), and aligned to the laser focus using a piezo-actuated precision stage.

\subsection{Parametric control}\label{subsec:param_control}
Direct mechanical interaction between the non-degenerate flexural modes is prevented by their large frequency spacing and narrow linewidths. Nevertheless, suitable modulation of radiation pressure backaction can enable controllable, light-mediated effective mechanical interactions~\cite{Weaver2017_SI,delpino2021nonhermitian_SI,Wanjura2023_SI}.
%

In the unresolved sideband regime or bad-cavity limit, where $\omega_j \ll \kappa$ as in our system, the cavity photon population $n_\text{c}$ responds to optical driving instantaneously on mechanical timescales. Moreover, with the control laser (Toptica CTL 1500) optimally tuned to the flank of the resonance $\Delta=\kappa/(2\sqrt{3})$, the dispersive coupling of mechanical displacement to the cavity frequency induces maximal modulations of $n_\text{c}$. Here, $\Delta=\omega_\text{l}-\omega_\text{c}$ denotes the detuning of the laser with frequency $\omega_\text{l}$ from the cavity resonance frequency. For a single resonator, the resulting mechanically modulated radiation pressure force acts back on the resonator's displacement, giving rise to the well-known optical spring effect~\cite{Aspelmeyer2014_SI},
%
where the resonator experiences a shift $\delta\omega_j = 2g_j^2\Delta/(\Delta^2 + \kappa^2/4)$ in its frequency. Here $g_j = \sqrt{\bar{n}_\text{c}} g_{0,j}$ is the optomechanical coupling enhanced by the average cavity population $\bar{n}_\text{c}$.

In our multimode system, the displacement of each resonator contributes a modulation to $n_\text{c}$. The ensuing modulated radiation pressure forces in principle generate `cross-resonator' optical springs, where the displacement of one resonator induces a force on the other ones~\cite{Shkarin2014_SI}.
%
For fixed drive laser intensity and non-degenerate modes, those forces cause off-resonant drives and do not induce appreciable interaction between the resonators. 
However, by temporally modulating the laser drive we can parametrically couple the modes, using cross-resonator optical springs to induce resonant interactions between mechanical modes. In particular, a harmonic modulation of the drive laser intensity with frequency $\omega_\text{m}$ set to the frequency difference $\omega_\text{m} = \omega_k - \omega_j$ (frequency sum $\omega_\text{m} = \omega_k + \omega_j$) of two modes $j$, $k$ stimulates a frequency conversion process and induces an effective mechanical beamsplitter (squeezing) interaction after adiabatic elimination of the cavity field~\cite{mathew2018synthetic_SI,delpino2021nonhermitian_SI}.
%
The depth $c_\text{m}$ of the harmonic modulation tone controls the effective interaction strength $|J_{jk}|$ ($|\lambda_{jk}|$) as
\begin{align}\label{eq:coupling_amplitudes}
    |J_{jk}|, |\lambda_{jk}| = c_\text{m} \frac{g_j g_k \Delta}{\Delta^2 + \kappa^2/4} = c_\text{m}\frac{\sqrt{\delta\omega_j\delta\omega_k}}{2},
\end{align}
while the phase offset $\phi_\text{m}$ of the modulation sets the interaction phase in the appropriate rotating frame as $J_{jk} \equiv |J_{jk}|e^{\i \phi_\text{m}}$ or $\lambda_{jk} \equiv |\lambda_{jk}|e^{\i \phi_\text{m}}$. This is crucial for controlling the coupling phase $\varphi$ discussed in the main text, and in particular, to generate the required imaginary couplings to construct the bosonic Kitaev chain (BKC) shown in Figs. 1 and 2. 

We note that while accurately estimating $g_j = \sqrt{\bar{n}_\text{c}} g_{0,j}$ can be challenging, the individual resonator spring shifts $\delta\omega_j$ are readily measured. Finally, modulation tones can be superimposed to induce multiple interactions, as all our mechanical frequencies are incommensurate. The required modulation tones are generated by an high-frequency lock-in amplifier (Zurich Instruments UHFLI) as well several additional signal generators (Siglent SDG1062X and SDG2122X) whose clocks are synchronized with that of the lock-in amplifier (section~\ref{sec:tone-referencing}). The signals are amplified and sent to fiber-coupled intensity modulators (Thorlabs LN81S-FC and Covega Mach-10 056) to modulate the control laser intensity.

\subsection{BKC Hamiltonian and dynamics}

The BKC Hamiltonian~\cite{McDonald2018_SI} 
%
that we implement combines beam-splitter and two-mode squeezing interactions between neighboring modes that are uniform along the chain, i.e.~$J_{jk}=J,\lambda_{jk}=\lambda$. It reads (see equation~(1) in the main text)
\begin{align}\label{eq:BKCHamiltonian}
    H_\text{BKC}= \sum_j \left( J a_{j+1}^\dagger a_j + \lambda a_{j+1}^\dagger a_j^\dagger + \mathrm{H.c} \right),
\end{align}
where mechanical modes are described by their field operators $a_j$ in a frame rotating along with the resonators.
For most of the main text, we focus on the case $J=\i |J|$ and $\lambda = \i |\lambda|$, in which case the Hamiltonian expressed in terms of the field quadratures $x_j\equiv(a_j+a_j^\dagger)/\sqrt{2}$, $p_j\equiv-\mathrm{i}(a_j-a_j^\dagger)/\sqrt{2}$ simplifies to~\cite{McDonald2018_SI}
%
\begin{align}\label{eq:BKCHamiltonianPi2}
    H_\text{BKC}= \sum_j (|\lambda| - |J|) x_{j+1}p_j + (|\lambda|+|J|) p_{j+1} x_j.
\end{align}

The field quadratures evolving according to the Hamiltonian~\eqref{eq:BKCHamiltonianPi2} obey Heisenberg-Langevin equations of motion
\begin{align}
    \dot x_j & = -\frac{\gamma_j}{2} x_j + (|J|+|\lambda|) x_{j-1} - (|J|-|\lambda|) x_{j+1} - \sqrt{\gamma_j} x_{j,\mathrm{in}} \label{eq:HL_eqs_1}\\
    \dot p_j & = -\frac{\gamma_j}{2} p_j + (|J|-|\lambda|) p_{j-1} - (|J|+|\lambda|) p_{j+1} - \sqrt{\gamma_j} p_{j,\mathrm{in}} \label{eq:HL_eqs_2}.
\end{align}
Here, $x_{j,\mathrm{in}}$ and $p_{j,\mathrm{in}}$ denote external driving quadratures, which quantify the probing forces that act on the resonators.

Equations~\eqref{eq:HL_eqs_1} and~\eqref{eq:HL_eqs_2} 
define the dynamical matrix $\mathcal{M}$
\begin{align}\label{eq:eomsXandP}
    \begin{pmatrix}
        \dot {\mathbf{x}} \\ \dot {\mathbf{p}}
    \end{pmatrix}
    & = 
    \begin{pmatrix} -\frac{\Gamma}{2} + \mathcal{H} & 0 \\ 0 & -\frac{\Gamma}{2} -\mathcal{H}^\mathrm{T} \end{pmatrix}
    \begin{pmatrix}
        \mathbf{x} \\ \textbf{p}
    \end{pmatrix}
    - \sqrt{\Gamma}
    \begin{pmatrix}
        \mathbf{x}_\mathrm{in} \\ \textbf{p}_\mathrm{in}
    \end{pmatrix} \notag \\
    & =
    \mathcal{M}
    \begin{pmatrix}
        \mathbf{x} \\ \textbf{p}
    \end{pmatrix}
    - \sqrt{\Gamma}
    \begin{pmatrix}
        \mathbf{x}_\mathrm{in} \\ \textbf{p}_\mathrm{in}
    \end{pmatrix}
\end{align}
with $\Gamma=\mathrm{diag}(\gamma_1,\dots,\gamma_N)$ and
in which we collected the quadratures in the vectors $\mathbf{x}\equiv(x_1,\dots,x_N)^\mathrm{T}$ and $\mathbf{p}\equiv(p_1,\dots,p_N)^\mathrm{T}$, and similarly for $\mathbf{x}_\mathrm{in}$ and $\textbf{p}_\mathrm{in}$.

We note the symmetry of the dynamical matrix which relates the dynamical matrices $M_x\equiv -\frac{\Gamma}{2} + \mathcal{H}$ and $M_p\equiv -\frac{\Gamma}{2} - \mathcal{H}^\mathrm{T} $ for the $x$ and $p$ quadratures, respectively.
For both open and periodic boundary conditions, the matrix $\mathcal{H}$ is defined as
$\mathcal{H}_{j,j+1} = |\lambda| - |J|$ and $\mathcal{H}_{j+1,j} = |\lambda| + |J|$, i.e., the effective coupling strengths to the left and to the right are asymmetric, $\mathcal{H}_{j,j+1}\neq \mathcal{H}_{j+1,j}$, so that the dynamic matrix $\mathcal{M}$ realises two copies of the Hatano-Nelson model.
For periodic boundary conditions, additionally, $\mathcal{H}_{N,1} = |\lambda| - |J|$ 
and $\mathcal{H}_{1,N} = |\lambda| + |J|$.

The susceptibility matrix encodes the steady-state system response to the quadratures of the external probe field at frequency $\omega$ and is calculated from the dynamical matrix $\mathcal{M}$ according to
\begin{align}
    \chi(\omega) & = (\mathrm{i}\omega\mathbb{1} + \mathcal{M})^{-1}.
\end{align}
As such, $\chi(\omega)$ describes the linear response of the quadratures $(\mathbf{x},\mathbf{p})^\mathrm{T}$ to a `drive' defined as $\mathbf{f}^{(\mathbf{q})}=\sqrt{\Gamma}(\mathbf{x}_\mathrm{in},\mathbf{p}_\mathrm{in})^\mathrm{T}$ with frequency $\omega$.

As we concern ourselves only with the mechanical response at resonance in this study, we introduce the short-hand $\chi \equiv \chi(0)$. In the main text, we refer to $\mathcal{M}$ also as the `non-Hermitian Hamiltonian matrix', which is thus the inverse of the resonant susceptibility $\chi=\mathcal{M}^{-1}$. We assume equal decay rates $\gamma_j=\gamma$ for all the mechanical modes, which holds in the experiment to good approximation since the damping rates are actively controlled via radiation pressure feedback signals derived from the measured displacements of all resonators. In the case of equal coupling amplitudes $|J|=|\lambda|=\mu$ (i.e. $J=\lambda=\iu \mu$), the resonant susceptibility then reads
\begin{align}\label{eq:chiExactEP}
\chi = -\frac{2}{\gamma}\begin{pmatrix}
1 \\
G & 1 & & & & &\text{\huge0} \\
G^2 & G & 1 \\
G^3 & G^2 & G & 1 \\
            & & & & 1 & -G & G^2 & -G^3 \\
            & & & & & 1 & -G & G^2 \\
            & \text{\huge0} & & & & & 1 & -G \\
            & & & & & &  & 1
\end{pmatrix},
\end{align}
where $G\equiv4\mu/\gamma$. This directly gives equation~(2) in the main text.

\subsection{Excitation and readout}
The displacement $z_j(t)$ of all resonators is imprinted on the intensity of the light of a second, far-detuned detection laser ($\Delta = -2.5\kappa$, Toptica CTL 1550) reflected off the cavity, which is then filtered spectrally from the modulated control laser using a tunable bandpass filter (DiCon) and detected by a fast photodiode (New Focus 1811, a.c.-coupled).
Displacement signals for the individual resonators are obtained by electronic frequency filtering using a lock-in amplifier (Zurich Instruments UHFLI). In particular, the detector signal is demodulated and low-pass filtered (third order, $3$-dB bandwidth $50$ kHz) in parallel using individual electronic local oscillators (LOs) for each resonator frequency to obtain records of their in-phase ($I_j$) and in-quadrature ($Q_j$) amplitudes. 

After appropriate normalization to the detector voltage level equivalent to each resonators' zero-point amplitude $x_\text{zpf}$ \cite{delpino2021nonhermitian_SI},
%
these components are formally equivalent to the resonator quadratures $x_j$ and $p_j$ in the rotating frame set by the electronic LOs. Importantly, as detailed in section~\ref{sec:tone-referencing}, we reference the control tones that establish our interactions to the same electronic LOs, such that their interaction phase in the rotating frame is deterministic~\cite{Wanjura2023_SI}.
%

To probe the susceptibility of the BKC to a harmonic force applied to one of its modes $j$, we modulate the radiation pressure of a third, resonant weak drive laser ($\Delta \approx 0$, Toptica CTL 1500) with modulation frequency $\omega_\text{d} = \omega_j$ and depth $c_\text{d}$. In the rotating frame, the resonator then feels the forces
\begin{align}
\begin{aligned}
    f_{x_j} &= \sqrt{\gamma}x_{j,\text{in}} = -\sin(\phi_\text{d}) c_\text{d} \frac{g_{0,j} \bar{n}_\text{d}}{2}, \\
    f_{p_j} &= \sqrt{\gamma}p_{j,\text{in}} = \cos(\phi_\text{d}) c_\text{d} \frac{g_{0,j} \bar{n}_\text{d}}{2},
\end{aligned}
\end{align}
driving its quadratures $x_j$ and $p_j$, respectively. Here, $\bar{n}_\text{d}$ is the average cavity photon population induced by the drive laser. We reference the drive tone again to the LO running in the lock-in amplifier for that particular resonator. Consequently, by setting the phase $\phi_d = 0, \pi/2$ of the (cosine) intensity modulation, the individual quadratures can be addressed. Finally, the full susceptibility matrix is reconstructed by driving each quadrature individually in a separate experiment and collecting the measured steady-state responses.

\subsection{Phase-referenced control tones}\label{sec:tone-referencing}
Even though the frequencies $\om^{(l)}$ of all control tones are distinct, the frequency relation $\om^{(l)} = \omega_j \pm \omega_k$ with the electronic LOs corresponding to resonators $j$ and $k$ lends each modulation tone $l$ a well-defined phase in the rotating frame. We label the tone $l = j\pm k$ by the resonator frequency combination $j\pm k$ it addresses.

To understand this phase relation, we absorb the explicit time dependence of each tone $\cos \left( \om^{(l)} t + \phim^{(l)} \right) = \cos \beta_l(t)$ into the instantaneous phase $\beta_l(t) = \om^{(l)} t + \phim^{(l)}$, and similarly for each LO labelled by its resonator index $j$. For convenience, we denote $\phim^{(j)}$ for the phase offset of LO $j$, even though it is not necessarily used as a modulation tone.
Phase coherence is now attained between the interaction tone $l$ and the \emph{combination} of both resonator LOs $j,k$, since the phase difference
\begin{align}
\begin{aligned}
\Delta \phim^{(j\pm k)} &= \beta_{j\pm k}(t) - \left( \beta_j(t) \pm \beta_k(t) \right) \\
&= \phim^{(j\pm k )} - \left( \phim^{(j)} \pm \phim^{(k)} \right) \label{met:eq:general-phase-coherence}
\end{aligned}
\end{align}
is stable and does not depend on the origin of time. Physically, this phase difference may be evaluated by generating a tone with the combined instantaneous phase $\beta_j(t) \pm \beta_k(t)$ through mixing of the LOs and subsequent high-pass (low-pass) filtering, and comparing that to the interaction tone.

As we work with relatively low modulation frequencies in the MHz range, we can take a digital approach to achieve phase coherence. In fact, the lock-in amplifier we use is a digital device based on $8$ numerical oscillators operating at a clock frequency of $1.8$~GHz, and synthesizes its output tones on-the-fly. We can therefore access the instantaneous phase $\beta_l(t)$ directly.
Note that the actual output signal lags the numerical oscillator by a time delay that depends on the oscillator index, equal to $16$ clock cycles per oscillator.

If only the LIA is used to generate control tones and no external signal generators are involved, we execute the following procedure. Before starting an experiment, at time $t_0$, we simultaneously access the instantaneous phases $\beta_l(t_0)$ of all numerical oscillators, generating LO and interaction tones alike. We then apply a phase shift equal to $-\beta_l(t_0)$ to all oscillators, effectively defining $t_0$ as the origin of time. Finally, we shift the interaction tone oscillators to generate the desired interaction phases $\arg(J_{jk}), \arg(\lambda_{jk})$ in the rotating frame. 

If an experiment requires more LOs and control signals than the $8$ numerical oscillators of the LIA can provide, we follow an extension of this procedure that is conceptually similar but involves a more elaborate phase bookkeeping. Suppose we want to monitor the dynamics of $N$ mechanical resonators under the influence of $M$ modulation tones, with $N+M>8$. As we require the LIA to analyse the mechanical signal, we implement up to $N=7$ resonator LOs using the lowest-index numerical oscillators. The remaining free LIA oscillators are used to generate the first $8-N$ interaction tones. The $M+N-8$ interaction tones that remain need to be generated by external signal generators.

During the phase referencing procedure, we temporarily use the LIA oscillator with highest index $m=8$ to `transfer' phase coherence between the LIA and the external signal sources. First, we achieve phase coherence between the oscillators internal to the LIA as described earlier. For each externally generated modulation tone $l = j \pm k$, we execute the following steps:
\setlist{nolistsep}
\begin{enumerate}[noitemsep]
    \item Set the frequency of LIA oscillator $m=8$ to match the frequency $\om^{(l)} = \omega_j \pm \omega_k$ of the external signal generator.
    \item Access the instantaneous phases $\beta_8(t)$ of the phase transfer oscillator $8$ and $\beta_j(t)$, $\beta_k(t)$ of the relevant resonator LOs.
    \item Use \eqref{met:eq:general-phase-coherence} to evaluate the current, \emph{arbitrary} phase $\Delta \phim^{(l)}$ of oscillator $8$ in the rotating frame.
    \item Shift the phase of oscillator $8$ by $-\Delta \phim^{(l)}$ to cancel its current phase offset. We have now achieved phase coherence between the tone generated by the phase transfer oscillator and the resonator LOs. \label{met:item:LO-PTO-coherent}
    \item Set the LIA to output a small voltage at $\om^{(l)}$ using oscillator $8$. This signal imprints a weak modulation on the control laser that we detect by tapping off part of the control laser light and feeding that onto a separate `monitor' detector.
    \item Analyse the monitor signal using the same LIA oscillator $8$ to evaluate the phase $\alpha_1$ of the imprinted modulation, combined with signal delay through part of the set-up.
    \item Disable the LIA output and enable a small output voltage at $\om^{(l)}$ on the external signal generator.
    \item Measure the phase $\alpha_2$ of the imprinted modulation using LIA oscillator $8$.
    \item Shift the phase of the external signal source by $-(\alpha_2 - \alpha_1)$ to achieve phase coherence between the LIA-generated tone and the externally generated tone.
    \item Shift the phase of the external signal source to generate the desired interaction phase $\arg(J_{jk}), \arg(\lambda_{jk})$ in the rotating frame. We are now done, and move on to the next external tone.
\end{enumerate} 
As a final step, the phase coherence of the interaction tone originally assigned to oscillator $8$ is restored by following the procedure above up to step~\ref{met:item:LO-PTO-coherent}.

\subsection{Dissipation control}

We implement measurement-based feedback to equalize the decay rates of the individual resonators to $\gamma_j=\gamma$, by applying a radiation pressure force proportional to each resonator's velocity. To do so, we modulate the weak drive laser with real-time feedback signals generated from the electronic displacement record by a digital signal processor (DSP, RedPitaya STEMlab 125-14) that implements a configurable bandpass filter for each resonator. Each filter's phase shift is tuned to provide velocity feedback to its respective resonator while the gain is varied to attain the desired damping rate. Details on calibration procedures can be found in~\cite{delpino2021nonhermitian_SI}.
%

\section{Supplementary theoretical details: Spectral winding and symmetry}

\subsection{Spectral winding in the BKC}
\label{sec:winding} 
As shown in Ref.~\cite{Wanjura2020_SI},
%
the BKC displays non-trivial non-Hermitian phases corresponding to directional amplification which occurs in opposite directions for the $x$ and the $p$ quadratures in which the respective end-to-end gain grows exponentially with the chain length $N$.

In the special case, $\varphi=\frac{\pi}{2}$, the situation is particularly simple. The equations for $x$ and $p$ decouple and we can consider the $x$ and $p$ chain separately. In particular, in the plane wave basis
$x_k \equiv \frac{1}{\sqrt{N}}\sum_j e^{\mathrm{i} k j} x_j$
and $p_k \equiv\frac{1}{\sqrt{N}}\sum_j e^{-\mathrm{i} k j} p_j$,
the dynamical matrix $\mathcal{M}_{xp}(k)$ governing the evolution of $\mathbf{q}_k=(x_k,p_k)^\mathrm{T}$ according to 
$\dot{\mathbf{q}}_k = \mathcal{M}_{xp}(k)\mathbf{q}_k - \sqrt{\gamma}\mathbf{q}_{k,\mathrm{in}}$ is already diagonal
\begin{align}\label{eq:PBCSpectrum}
    \mathcal{M}_{xp}(k)
    =
    \begin{pmatrix}
        \omega_{x}(k) & 0 \\
        0 & \omega_{p}(k)
    \end{pmatrix},
\end{align}
with the spectrum $\omega_{x,p}(k)=-\mathrm{i}\frac{\gamma}{2}-2J\sin k \mp 2\mathrm{i}\lambda\cos k$ given in Eq.~(3) of the main text.

Accordingly, we define the winding numbers $\nu_{x,p}$ on the spectrum for periodic boundary conditions (PBC)~\eqref{eq:PBCSpectrum} $\omega_{x,p}(k)$ as given in Eq.~(3) of the main text through~\cite{Wanjura2020_SI}
%
\begin{align}\label{eq:winding}
    \nu_{x,p} \equiv \frac{1}{2\pi\mathrm{i}} \int_0^{2\pi} \mathrm{d}k \, \frac{\partial \omega_{x,p}(k)/\partial k}{\omega_{x,p}(k)},
\end{align}
which 
are $\nu_x=-1$ and $\nu_p=+1$ for $2|\lambda|>\gamma/2$ and $\nu_x=\nu_p=0$ otherwise.

The Hamiltonian of the generalized BKC is given by
\begin{align}
    H_\text{BKC} & = \sum_j\left[
        |J| e^{\mathrm{i}\varphi} a_{j+1}^\dagger a_j + |\lambda| e^{\iu \varphi} a_{j+1}^\dagger a_j^\dagger + \mathrm{H.c.}
    \right].
\end{align}
In general, this Hamiltonian leads to coupled equations of motion for $x$ and $p$, so the formula above, Eq.~\eqref{eq:winding}, cannot be applied directly to compute the winding numbers. Instead, we will diagonalise the Bloch dynamical matrix under periodic boundary conditions to find a transformed set of quadratures for which the dynamical equations decouple.

To study the system under periodic boundary conditions, we introduce $a_k\equiv\frac{1}{\sqrt{N}}\sum_j e^{\iu k j} a_j$, $a_k^\dagger\equiv\frac{1}{\sqrt{N}}\sum_j e^{-\iu k j} a_j^\dagger$ and $a_{-k}\equiv\frac{1}{\sqrt{N}}\sum_j e^{-\iu k j} a_j$, $a_{-k}^\dagger\equiv\frac{1}{\sqrt{N}}\sum_j e^{\iu k j} a_j^\dagger$. This yields
\begin{align}
    H_\text{BKC}^\mathrm{PBC} & = \sum_j\left[
        2 |J| \cos(k+\varphi) a_k^\dagger a_k +|\lambda| (e^{\mathrm{i}k} e^{\iu \varphi} a_k^\dagger a_{-k}^\dagger + e^{-\mathrm{i}k}e^{-\iu \varphi} a_k a_{-k})
    \right].
\end{align}
This allows us to compute the dynamical matrix under periodic boundary conditions for the vector 
$\alpha_k^T=(a_k, a_{-k}^\dagger)^\mathrm{T}$ with $k\in[0,2\pi)$ satisfying $\dot \alpha_k = \mathcal{M}_\alpha \alpha_k - \sqrt{\gamma}\alpha_{k,\mathrm{in}}$ with
\begin{align}\label{eq:BlochDynMatGeneral}
    \mathcal{M}_\alpha(k) & = \begin{pmatrix}
    -\mathrm{i}\frac{\gamma}{2} + 2 |J| \cos(k+\varphi) & 2e^{\iu\varphi}|\lambda|\cos k \\
    -2e^{-\iu\varphi}|\lambda|\cos k & -\mathrm{i}\frac{\gamma}{2} - 2 |J| \cos(-k+\varphi)
    \end{pmatrix}.
\end{align}
The eigenvalues of this matrix are given by
\begin{align}\label{eq:eigenvaluesPhi}
    \omega_\pm(k) &= -\mathrm{i}\frac{\gamma}{2} - \left(2|J|\sin\varphi\sin k \pm 2\iu \sqrt{|\lambda|^2 - |J|^2 \cos^2\varphi}\cos k \right).
\end{align}
At $\varphi=\frac{\pi}{2}$, we recover $\omega_+(k)=\omega_x(k)$ and $\omega_-(k)=\omega_p(k)$ of Eq.~(3) in the main text.
As we show below, importantly, the corresponding eigenvectors are independent of $k$, which implies that there exists a local transformation that maps back to the BKC at $\varphi=\frac{\pi}{2}$~\cite{McDonald2018_SI}.
%
As a consequence the correspondence between non-trivial, non-Hermitian topology and directional end-to-end amplification~\cite{Wanjura2020_SI} 
%
applies to the BKC for all $\varphi$.
If the transformation was not local, there could, in principle, still be a basis in which the system displays directional amplification, but the amplification does not have to be end-to-end or even directional in real space.
Since here the transformation is local, it is justified to study the winding of $\omega_\pm(k)$ which can then again be related to directional end-to-end amplification under open boundary conditions. 
Analogously, the winding numbers $\nu_\pm$ of $\omega_\pm(k)$ are defined according to
\begin{align}\label{eq:windingGeneral}
    \nu_\pm & \equiv \frac{1}{2\pi\mathrm{i}} \int_0^{2\pi} \mathrm{d}k \, \frac{\partial \omega_{\pm}(k)/\partial k}{\omega_{\pm}(k)}.
\end{align}

We notice from the expression for the spectrum~\eqref{eq:eigenvaluesPhi} there are two cases: either ${\left\lvert (J/\lambda)\cos\varphi\right\rvert<1}$ or ${\left\lvert (J/\lambda)\cos\varphi\right\rvert>1}$.
In the first case, the term under the square root in Eq.~\eqref{eq:eigenvaluesPhi} is positive, so $\omega_\pm(k)$ is complex and the point gap is open (see Fig.~3 in the main text). The winding number~\eqref{eq:windingGeneral} can be $+1$, $-1$ or $0$.
In the second case, the term under the square root is negative, so the $k$-dependent terms of the spectrum are all real and the point gap is closed (see Fig.~3 in the main text).
The winding number~\eqref{eq:windingGeneral} is always zero.

Next, we show explicitly that the eigenvectors are independent of $k$.
Introducing, $y\equiv |J/\lambda|\cos\varphi$, we can express the corresponding eigenvectors as
\begin{align}\label{eq:eigenvectorsGeneral}
    \ket{\Psi_+} \equiv & \frac{1}{\mathcal{N}_+}(e^{\iu\varphi}[-y+\sqrt{y^2-1}],1)^\mathrm{T} \notag, \\
    \ket{\Psi_-} \equiv & \frac{1}{\mathcal{N}_-}(e^{\iu\varphi}[-y-\sqrt{y^2-1}],1)^\mathrm{T},
\end{align}
with the normalization constant
\begin{align}
    \mathcal{N}_\pm & \equiv \begin{cases}
        \sqrt{1+(y\pm\sqrt{y^2-1})^2} & : \lvert y\rvert\geq 1 \\
        \sqrt{2} & : \lvert y\rvert < 1
    \end{cases}.
\end{align}
The eigenvectors also differ in the two cases mentioned above. 
In the first case ${\left\lvert (J/\lambda)\cos\varphi\right\rvert<1}$ (point gap open), defining $y\equiv\left\lvert J/\lambda\right\rvert \cos\varphi=\cos\eta$ simplifies the eigenvectors to
\begin{align}
    \ket{\Psi_+} \equiv & \frac{1}{\sqrt{2}}(-e^{\iu(\varphi-\eta)},1)^\mathrm{T}, \quad\quad\quad
    \ket{\Psi_-} \equiv \frac{1}{\sqrt{2}}(-e^{\iu(\varphi+\eta)},1)^\mathrm{T}.
\end{align}
In the second case ${\left\lvert (J/\lambda)\cos\varphi\right\rvert>1}$ (point gap closed), defining $\cosh\xi\equiv y$ leads to the eigenvectors
\begin{align}
    \ket{\Psi_+} \equiv & \frac{1}{\sqrt{1+e^{-2\lvert\xi\rvert}}}(-e^{-\lvert\xi\rvert},1)^\mathrm{T}, \quad\quad\quad
    \ket{\Psi_-} \equiv \frac{1}{\sqrt{1+e^{2\lvert\xi\rvert}}}(-e^{\lvert\xi\rvert},1)^\mathrm{T}.
\end{align}

\subsection{Parity symmetry of the BKC}

We mentioned in the main text that the BKC preserves parity symmetry~\cite{Kawabata2019symmetry_SI} 
%
for any phase $\varphi$.
Intuitively, parity symmetry implies that reflecting the system and performing a suitable \emph{local} transformation, the system remains unchanged. In the BKC at $\varphi=\frac{\pi}{2}$ this is particularly easy to see since reflecting the system and swapping $x$ and $p$ we obtain the same response (see the susceptibility matrices in Fig.~1).
This can be generalized to arbitrary $\varphi$.
In particular, we notice that the eigenvalues in Eq.~\eqref{eq:eigenvaluesPhi} satisfy
\begin{align}
    \omega_+\left(k+\frac{\pi}{2}\right) = \omega_-\left(-k+\frac{\pi}{2}\right).
\end{align}
This leads to a condition for $\mathcal{M}_\alpha(k)$
\begin{align}\label{eq:paritySymmetry}
    \mathcal{M}_\alpha\left(k+\frac{\pi}{2}\right) V
    = V \mathcal{M}_\alpha\left(-k+\frac{\pi}{2}\right),
\end{align}
with $V=(\ket{\Psi_+},\ket{\Psi_-})$.
Eq.~\eqref{eq:paritySymmetry} is the formal definition of parity symmetry~\cite{Kawabata2019symmetry_SI}.
%
Note that as a consequence of parity symmetry, the two winding numbers $\nu_\pm$, Eq.~\eqref{eq:windingGeneral}, associated with $\omega_\pm(k)$ have opposite sign.
Since the eigenvectors~\eqref{eq:eigenvectorsGeneral} are independent of $k$, the transformation $V$ is local.
Alternatively, we can take a different view point: as long as Eq.~\eqref{eq:paritySymmetry} holds and $V$ is local ($k$ independent), we can compute the winding numbers of $\omega_\pm(k)$ and we can relate this winding number to directional amplification under open boundary conditions. In that sense, parity symmetry \emph{protects} non-Hermitian topology and the associated scattering properties.

This symmetry is broken in the presence of detuning since in that case the eigenvectors become $k$ dependent and the transformation $V$ is no longer local.

\section{Thermal steady-states}\label{thermalsteadystates}
In the absence of coherent driving, the BKC is subject to thermal forces. Squeezing in the chain then provides amplification of the resulting thermal fluctuations above their equilibrium level in the absence of interactions, as shown in Fig.~2. 

For a BKC with four sites coupled to individual thermal baths with occupancies $n_{j,\text{th}}$, we calculate the average phonon populations $n_j^\text{OBC}$ and $n_j^\text{PBC}$ under open and closed boundary conditions, respectively. For simplicity, we assume $n_{j,\text{th}} = n_\text{th}$, which holds to reasonable approximation in our experiment due the provided feedback cooling. Using the quantum regression theorem~\cite{Meystre2007elements_SI,delpino2021nonhermitian_SI}, 
%
the populations of the first resonator are then given by
\begin{align}
n_1^\text{OBC} &= n_\text{th} \left( 1 + G^2/4 + 3G^4/16 + 5 G^6 /32 \right), \label{eq:bkc-thermpop-open} \\
n_1^\text{PBC} &= n_\text{th} \left( 1 + \frac{1}{2} \frac{G^2}{1-G^2} \right) \label{eq:bkc-thermpop-closed}.
\end{align}
These expressions define the theory lines in Fig.~2b. While equation \eqref{eq:bkc-thermpop-open} remains finite for any value of $G$, the dynamical instability under periodic boundary conditions is signified by the fact that $n_1^\text{PBC}$ tends to infinity for $G \to 1$ in equation \eqref{eq:bkc-thermpop-closed}.

Since the closed chain with equal bath occupations is translationally invariant, expression \eqref{eq:bkc-thermpop-closed} holds in fact for all resonators. This does not apply to the open chain, where we find that the two outer sites $n_1^\text{OBC} = n_4^\text{OBC}$ have equal populations, whereas the inner sites have
\begin{align}
 n_2^\text{OBC} = n_3^\text{OBC} =  n_\text{th} \left( 1 + G^2/2 + 3G^4/16 \right),
 \end{align}
 that grow as a polynomial in $G$ with lower degree than in \eqref{eq:bkc-thermpop-closed}. This reflects the shorter length of the effective amplification chains that feed the inner sites.

\section{Exponentially enhanced responsivity for non-trivial non-Hermitian topology}

Here, we show that the exponential increase in responsivity reported in the main text is a signature of non-trivial, non-Hermitian topology.
An exponential sensitivity to boundary conditions in non-Hermitian systems with a point gap was remarked on in Ref.~\cite{Kunst2018Biorthogonal_SI} 
%
and explored for sensing applications in Ref.~\cite{Budich2020_SI}.
%
The setup proposed by McDonald et al.~\cite{McDonald2020_SI} 
%
which we explore here is instead based on the exponential amplification of the unperturbed system under open boundary conditions and has the advantage that the system cannot become dynamically unstable for some values of the quantity that is to be sensed.
Here, we identify the origin of the complex plane as a relevant reference point and rigorously prove that non-trivial values of a winding number calculated w.r.t.~the origin (under PBC) are in one-to-one correspondence with the exponential scaling of the sensitivity with system size (under OBC).

In the sensing setup considered in the main text, we consider the BKC~\eqref{eq:BKCHamiltonian} at $\varphi=\pi/2$ and couple the $N$th site of the chain to a system which we would like to sense and we assume that we introduce a detuning on the last site.
Without this detuning, the equations of motion can be written in the block form of $\mathcal{M}$ in Eq.~\eqref{eq:eomsXandP}. In this configuration, the $x$ chain is amplified from $1$ to $N$ and the $p$ chain in the opposite direction~\cite{McDonald2020_SI}.
%

The introduction of the detuning $\varepsilon$ on the last site couples the $x$ and $p$ quadratures on that site
\begin{align}
    \dot x_N & = -\varepsilon p_N + \left(-\frac{\Gamma}{2}\mathbf{x} + \mathcal{H} \mathbf{x}\right)_N - \sqrt{\gamma} x_{N,\mathrm{in}}, \\
    \dot p_N & = \varepsilon x_N + \left(-\frac{\Gamma}{2} \mathbf{p}- \mathcal{H}^\mathrm{T} \mathbf{p}\right)_N - \sqrt{\gamma} p_{N,\mathrm{in}}.
\end{align}
This coupling allows for a signal in the $x$ quadrature to be amplified along the $x$ chain from site $1$ to site $N$ with a gain that is exponential in the chain length, converted to a signal in $p$ depending on the strength of the detuning $\varepsilon$, and amplified back to site $1$ with exponentially large gain~\cite{McDonald2020_SI}.
%

We showed in Ref.~\cite{Wanjura2020_SI}
%
that the gain is exponential in system size if and only if the system's non-Hermitian topology is non-trivial. As the conversion from $x$ to $p$ quadrature depends on the strength of the detuning at the $N$th site, it is plausible that the sensitivity with which we can sense the change in detuning scales exponentially with system size.

We now proceed to make this argument more rigorous.
Adopting the Dirac notation for matrix elements, we can write the dynamical matrix $\mathcal{M}^{(\varepsilon)}$ of the system that includes the detuning at the last site
\begin{align}\label{eq:dynMatDetuning}
    \mathcal{M}^{(\varepsilon)}
    & =
    \begin{pmatrix} -\frac{\Gamma}{2} + \mathcal{H} & -\varepsilon\ketbra{N}{N} \\[1ex] \varepsilon\ketbra{N}{N} & -\frac{\Gamma}{2} - \mathcal{H}^\mathrm{T} \end{pmatrix}.
\end{align}
We now derive an analytic formula for the susceptibility matrix $\chi^{(\varepsilon)} = (\mathrm{i}\omega\mathbb{1} + \mathcal{M}^{(\varepsilon)})^{-1}$ and relate it to the susceptibility matrix without detuning $\chi(\omega)$
\begin{align}\label{eq:susceptbilityMatrixDisorderless}
    \chi(\omega) & = \begin{pmatrix} (\mathrm{i}\omega\mathbb{1} -\frac{\Gamma}{2} + \mathcal{H})^{-1} & 0 \\ 0 & (\mathrm{i}\omega\mathbb{1} -\frac{\Gamma}{2} - \mathcal{H}^\mathrm{T})^{-1} \end{pmatrix}
    = \begin{pmatrix} \chi_{xx}(\omega) & 0 \\ 0 & \chi_{pp}(\omega) \end{pmatrix}.
\end{align}
To that purpose, we employ a formula for the inverse~\cite{schott2016matrix_SI} 
%
of a sum of a matrix $A$ and a rank-one matrix $E$ and apply it recursively---a technique that was also used in~\cite{Wanjura2020_SI} 
%
to derive the one-to-one correspondence between non-trivial, non-Hermitian topology and directional amplification.
Concretely, $(A+E_j)^{-1} = A^{-1} - \frac{1}{1+g_j} A^{-1} E_j A^{-1}$ with $g_j\equiv \mathrm{tr}(A^{-1}E_j)$~\cite{Miller1981On_SI}.
%

Here, we need to calculate
$\chi^{(\varepsilon)}(\omega)=(\mathrm{i}\omega\mathbb{1} + \mathcal{M} + \varepsilon [\ketbra{N,p}{N,x} - \ketbra{N,x}{N,p}])^{-1}$.
Both $\ketbra{N,p}{N,x}$ and $\ketbra{N,x}{N,p}$ are rank-one, so we can apply the formula above recursively.
This gives rise to the following expression for $\chi^{(\varepsilon)}(\omega)$
\begin{align}\label{eq:susceptibilitySensing}
    \chi^{(\varepsilon)}(\omega) =
    \chi
    + \frac{\varepsilon}{1+\varepsilon^2\left(\chi_{x_N \to x_N}\right)^2}
    & \left(
    \chi \ketbra{p,N}{x,N} \chi
    - \chi \ketbra{x,N}{p,N} \chi
    \right)
    + \varepsilon
    \left(
    \chi \ketbra{x,N}{x,N} \chi
    - \chi \ketbra{p,N}{p,N} \chi
    \right).
\end{align}
For clarity, we omitted the argument $\omega$ on the right-hand-side of Eq.~\eqref{eq:susceptibilitySensing}.
We obtain the optimal sensitivity to changes in $\varepsilon$ by measuring the response at the first site, i.e.,
\begin{align}\label{eq:susceptibilityX1P1Sensing}
    \chi^{(\varepsilon)}_{p_1\to x_1} & = - \chi^{(\varepsilon)}_{x_1\to p_1} = \frac{\varepsilon}{1+\varepsilon^2\left(\chi_{x_N \to x_N}\right)^2}
    \left(\chi_{x_1\to x_N}\right)^2
\end{align}
in which $\chi_{x_1\to x_N}\propto e^{\alpha N}$ with $\alpha>1$ in topologically non-trivial phases.
Eq.~\eqref{eq:susceptibilityX1P1Sensing} is exact for the BKC at $|J|=|\lambda|$ and $\varphi=\pi/2$ which constitutes the exceptional point at which the reverse transmission is exactly zero leading to the cancellation of contributions from the third term in Eq.~\eqref{eq:susceptibilitySensing} to $\chi^{(\varepsilon)}_{p_1\to x_1}$.
Otherwise, Eq.~\eqref{eq:susceptibilityX1P1Sensing} is exact up to an exponentially small correction of the order $e^{-\alpha N}$ with $\alpha>1$.

Close to $\varepsilon=0$, $\chi^{(\varepsilon)}_{p_1\to x_1}$ is linear in $\varepsilon$ with a slope that grows exponentially in system size. This is straightforward to see from Eq.~\eqref{eq:susceptibilityX1P1Sensing}
\begin{align}
    \frac{\partial \chi^{(\varepsilon)}_{p_1\to x_1}}{\partial\varepsilon} & = - \frac{\partial \chi^{(\varepsilon)}_{x_1\to p_1}}{\partial \varepsilon} 
    = \frac{1-\varepsilon^2 \left(\chi_{x_N \to x_N}\right)^2}{\left[1+\varepsilon^2\left(\chi_{x_N \to x_N}\right)^2\right]^2}
    \left(\chi_{x_1\to x_N}\right)^2.
\end{align}
Around $\varepsilon = 0$, we define the linear responsivity as given in Eq.~(4) of the main text
\begin{align}
    \mathcal{R} \equiv \gamma \left\lvert\frac{\partial \chi^{(\varepsilon)}_{x_1\to p_1}}{\partial\varepsilon}
    \right\rvert_{\varepsilon=0} = \gamma \left(\chi_{x_1 \to x_N}\right)^2.
\end{align}

At the exceptional point, i.e.~$|J|=|\lambda|$ and $\varphi=\pi/2$, the susceptibility matrix can be calculated exactly analytically, see Eq.~\eqref{eq:chiExactEP}. 
With this, we find that the responsivity at $\varepsilon=0$ is simply given by
\begin{align}
    \mathcal{R} = \frac{4}{\gamma} G^{2(N-1)}
    = \frac{4}{\gamma} \left(4\frac{|\lambda|}{\gamma}\right)^{2(N-1)}.
\end{align}

\section{Effects of optomechanical nonlinearity}\label{sec:optomechanical-nonlinearity}

In our system, the detuning $g_{0,j} z_j$ due to a mechanical displacement $z_j$ (expressed in units of the zero-point amplitude $x_\text{zpf}$, i.e.~$z=x/x_{\text{zpf}}$) of a resonator $j$ might be similar in magnitude to the cavity linewidth $\kappa$. In that case, the linear approximation for the Lorentzian cavity response at control laser detuning $\Delta = \kappa/(2\sqrt{3})$, which underlies the optical spring effect causing a frequency shift and facilitating interactions, becomes inadequate. This occurs generally for relatively large coherent driving, but in our experiments particularly when vibrations are amplified to significant amplitude due to parametric amplification. It leads to deviations from linear theory that we observe in the main text. Specifically, they lead to:
\begin{itemize}
    \item Amplitude saturation in the self-oscillation regime of main text Fig. 2, in the case of the BKC with open boundary conditions. In this regime, linear response theory breaks as it predicts a single steady state that is both dynamically and thermodynamically unstable, characterized by fluctuation eigenvalues having positive imaginary parts, and diverging average thermal energy. The figure illustrates the stabilization of the solution, a phenomenon consistent with dispersive nonlinearities and gain saturation.
    \item Reduced responsivity w.r.t. to the linear prediction in main text Fig. 4, especially noticeable for largest chain lengths, which exhibit the most amplification. This decrease in responsivity can be attributed to dispersive shifts and gain saturation.
\end{itemize}

In this section we discuss the basic effects of such nonlinearity, demonstrating that including higher-order terms in the optical force leads to an optically-induced effective nonlinearity in the mechanical resonators' equation of motion. The dominant effects observed align with those previously described.

\subsection{Single resonator}
Consider a single resonator dispersively coupled to an optical cavity. In the bad cavity limit, the  cavity photon population induced by a drive laser with average detuning $\Delta$ adjusts (i.e., reaches its steady-state) instantaneously with position, and is given by
\begin{equation}
    \bar{n}_\text{c}(z) = h(2[\Delta + g_0 z]/\kappa) n_\text{max},
\end{equation}
where $n_\text{max}$ is the maximum population at resonance ($\Delta, z = 0$), and
\begin{align}
    h(u) = \frac{1}{1+u^2} \label{eq:lorentzian-resp},
\end{align}
with $u = 2\Delta / \kappa$ the nondimensionalized Lorentzian cavity response function. Note that mechanical motion that is nonlinearly transduced onto the intracavity intensity generates harmonics of the mechanical oscillation in the cavity field, i.e. in the frequency spectrum of $\bar{n}_\text{c}$~\cite{leijssen2017nonlinear_SI}. 
%
Likewise, the magnitude of the fundamental harmonic at the mechanical frequency can change non-linearly with its amplitude. The cavity field exerts a force $F_c(z)= \hbar g_0 \bar{n}_\text{c}(z) / x_\text{zpf}$ on the resonator, causing dynamical backaction. The fact that $\bar{n}_\text{c}(z)$ is not a linear function of $z$ means that also the dynamical backaction has a nonlinear character. We estimate its magnitude below.
Under the backaction force  $f_c(z)$, the equation of motion for $z$ reads
\begin{equation}
    \ddot{z}(t) = -\omega^2 z(t) - \gamma \dot{z}(t) + f_c(z),
\end{equation}
with  $f_c(z) = 2\omega_0 g_0n_\mathrm{max} h(u)$.

The standard, linear optical spring effect is obtained by expanding $f_c$ linearly in $z$. As our strong control laser is detuned to induce the maximum optical spring shift ($\Delta = \kappa/(2\sqrt{3})$), the second derivative $h^{\prime\prime}(1/\sqrt{3})$ of the cavity response is zero, such that next higher order order term in the expansion of $f_c$ is cubic in $z$. This leads to an effective resonator's equation of motion
\begin{align}
    \ddot{z}(t)  \approx  -\omega^2 z(t) - \gamma \dot{z}(t) - \alpha z(t)^3,
\end{align}
with a Duffing nonlinear coefficient 
\begin{align}
    \alpha = \frac{8 \omega g_0^4 n_\text{max} h^{\prime\prime\prime}(2\Delta/\kappa)}{3\kappa^3} = -6 \omega \delta\omega \frac{g_0^2}{\kappa^2},
\end{align}
with $\delta\omega=2g_0^2\bar{n}_\text{c}(z=0)\Delta/(\Delta^2+\kappa^2/4)$ and where the last expression is valid at the maximum spring shift detuning $\Delta = \pm \kappa/(2\sqrt{3})$ (corresponding to $u=\pm 1/\sqrt{3}$).

For small oscillation amplitudes $A$, the approximate effect of such a Duffing nonlinearity is to induce a detuning, i.e. a change of the mechanical frequency. In other words, the effect can be perceived as an effective, mechanical amplitude-dependent correction to the optical spring shift. The shifted resonance frequency $\mu$ can be found using the Harmonic Balance method~\cite{Krack_2019_SI} 
%
to be
\begin{align}
    \mu^2 = \omega^2 + \frac{3 \alpha}{4} A^2 \approx (\omega + \delta\omega_\text{NL})^2,
\end{align}
where $\omega$ already includes the linear spring shift $\delta\omega \ll \omega$. The nonlinear contribution to frequency shift then reads approximately
\begin{align}
    \delta\omega_\text{NL} = -\delta\omega \frac{9g_0^2}{4\kappa^2}A^2 \label{eq:delta-nl}
\end{align}
and counteracts the linear optical spring. Note that the factor $g_0 A / \kappa$ appearing in \eqref{eq:delta-nl} expresses the extent of the Lorentzian cavity response function \eqref{eq:lorentzian-resp} that is explored by the mechanical oscillations.

\subsection{Many resonators}
As we saw above, the optomechanical nonlinearity can cause an effective correction to the optical spring effect. In multi-resonator systems, in particular when the resonators are coupled to each other through a modulated spring effect, the nonlinear effects can have various consequences. On the one hand, the effective mechanical frequency shift introduced due to nonlinearity creates an effective detuning of the affected resonator(s). On the other hand, the effective amplitudes of the couplings can change due to nonlinearity.

The displacements of multiple mechanical resonators, parametrized by $z_i=x_i/x_{\mathrm{zpf},i}$, cause cavity frequency shifts that modulate the intracavity intensity.  By applying suitable parametric driving, this effect generates BS and TMS interactions through a linear cross-spring effect, as outlined in Sec.~\ref{subsec:param_control} and~\cite{delpino2021nonhermitian_SI}. 
%
We extend the analysis of nonlinear effects in a single resonator to show that crossed nonlinearities in a resonator chain are expected under similar conditions and list the dominant contributions enhanced by modulations of the intracavity intensity. The optical force for $N$ resonators reads
    \begin{equation}
        f_{c,i}(\textbf{z})=2\omega_in_\mathrm{max} g_{0,i}n_\mathrm{max}h(u(\textbf{z})),
    \end{equation}
    where we assumed the global bad-cavity limit $\omega_i\ll\kappa$, we used the shorthand $\textbf{z}=(z_1,z_2,\cdots,z_N)^T$ and made explicit that the cavity shift depends on the amplitude of all resonators $u(\textbf{z})=u_{0}+2\sum_{i}g_{0,i}z_{i}/\kappa$. We now apply a perturbative expansion in the displacements $z_i\ll\kappa/g_{0,i}$, analog to that of a single resonator. In this expansion, the constant term $f_{c,i}^{(0)}=\hbar G_in_\mathrm{max}h(u_0)$ causes a shift in the resonator's equilibrium position. The first order linear correction reads
    \begin{align}\label{eq:linear_force_many}
        f_{c,i}^{(1)}=2\omega_i g_{0,i}n_\mathrm{max}\textbf{z}^T\nabla_{\textbf{z}}u(\textbf{z})=-\frac{4\omega_i g_{0,i}n_\mathrm{max}}{\kappa}\sum_j g_{0,j}z_j=0.
    \end{align}
    where $\nabla_{\textbf{z}}$ denotes the gradient vector with respect to the $\textbf{z}$ coordinates.
    This correction recovers both self and crossed spring shift linear effects. Namely, Eq. \eqref{eq:linear_force_many} accounts for a force exerted on resonator $z_i$ due to a cavity modulation enacted by the oscillation of resonator $z_j$. This force is time-dependent, and may become resonant with the resonator $i$ if the intracavity intensity is modulated in time: $n_\mathrm{max}c_m\mapsto n_\mathrm{max}(t)=n_\mathrm{max}\cos(\omega_\text{m}t)$. The force is equivalent to a time-dependent interaction bilinear Hamiltonian (up to a constant term)
    \begin{equation}
        H_{\mathrm{int}}^{(1)}=-\sum_{i}\int dz_{i}f_{c,i}^{(1)}=\sum_{i,j\neq i}\frac{t_{i,j}}{\omega_i}z_{i}z_{j},
    \end{equation}
    where $t_{i,j}=2c_m\sqrt{3}\omega_ig_{0,i}g_{0,j}\bar{n}_{c}/\kappa=2c_m\omega_i\sqrt{\delta\omega_i\delta\omega_j}$.  Difference frequency modulations with $\omega_m=\omega_i-\omega_j$ (sum frequency modulation $\omega_i+\omega_j$) lead to the hopping and two-mode squeezing interactions described in the Sec. \ref{subsec:param_control}, with amplitudes $|J_{i,j}|,|\lambda_{i,j}|=t_{i,j}/(4\omega_i)$, recovering equations  Eqs. \eqref{eq:coupling_amplitudes}.
     The 2nd order term reads $f_{c,i}^{(2)}=\omega_ig_{0,i}n_{r}\textbf{z}^{T}\mathcal{H}(\textbf{z})\textbf{z}$ where  $\mathcal{H}_{i,j}(\textbf{z})=\partial^2 h(u(\textbf{z})/(\partial z_i \partial z_j)$ stands for the Hessian matrix of the function $h(u(\textbf{z}))$. This second order contribution again vanishes at the maximum spring shift operation point $\Delta=\Delta_\pm$ ($u_0=u_\pm\equiv\pm1/\sqrt{3}$):
     \begin{equation}
         f_{c,i}^{(2)}=\frac{8\omega_in_{r}g_{0,i}\left(3u_{0}^{2}-1\right)}{\kappa^{2}\left(1+3u_{0}^{2}\right)^{3}}\sum_{j,k}g_{0,j}g_{0,i}z_{j}z_{k},
     \end{equation}
     The third order contribution can be written succintly as $f_{c,i}^{(3)}=\frac{\omega_ig_{0,i}n_{r}}{3}\sum_{j,k,l}\frac{\partial^{3}h(u(\mathbf{z}))}{\partial z_{j}\partial z_{k}\partial z_{l}}z_{j}z_{k}z_{l}$ and can be readily evaluated to be
    \begin{align}
             f_{c,i}^{(3)}=&-\frac{256u_{0}\left(u_{0}^{2}-1\right)\omega_i\bar{n}_{c}g_{0,i}}{\kappa^{3}\left(u_{0}^{2}+1\right)^{4}}\sum_{j,k,l}g_{0,k}g_{0,k}g_{0,l}z_{j}z_{k}z_{l}.
         \end{align}
    At $u_0=u_\pm$, the equation of motion for the amplitude $z_i$ then reads
        \begin{equation}\label{eq:mean_field}
        \ddot{z}_{i}\approx-\Omega'^{2}z_{i}-\gamma\dot{z}_{i}-\sum_{j\neq i}t_{i,j}z_{j}-\sum_{j,k,l\neq i}\alpha_{i,j,k,l}z_{j}z_{k}z_{l},
        \end{equation}
    with $\alpha_{i,j,k,l}=\mp 6\sqrt{3}\omega_i\bar{n}_cg_{0,i}g_{0,j}g_{0,k}g_{0,l}/\kappa^{3}$.
    
    The nonlinear contributions to this equation can be derived from a Hamiltonian
    \begin{equation}\label{eq:quartic_potential}
    H_{\text{int}}^{(3)}=\frac{1}{4}\sum_{i,j,k,l}\alpha_{i,j,k,l}z_{i}z_{j}z_{k}z_{l}.
    \end{equation}
    The experiment focuses on the slow evolution of mechanical resonators rotating at frequency $\omega=\omega_j$. This allows substantially simplifying Eq.~\eqref{eq:3orderH} by filtering out the fast dynamics via a rotating wave approximation (RWA). To apply it, we express Eq.~\eqref{eq:quartic_potential} in terms of canonical operators in a rotating frame at frequencies $\omega_j$, leading to
    \begin{equation}\label{eq:3orderH}
        H_{\text{int}}^{(3)}=\frac{1}{16}\sum_{i,j,k,l}\alpha_{ijkl}(t)(a_{i}\zeta_i+a_{i}^{\dagger}\zeta_i^*)(a_{j}\zeta_j+a_{j}^{\dagger}\zeta_j^*)(a_{k}\zeta_k+a_{k}^{\dagger}\zeta_k^*)(a_{l}\zeta_l+a_{l}^{\dagger}\zeta_l^*),
    \end{equation}
    with $\zeta_j=e^{i\omega_j t}$.  Fast-evolving contributions in the rotating frame can be averaged out, as they are assumed to only introduce perturbative corrections corrections to the dynamics over periods $T_j=2\pi/\omega_j$. Noting that the nonlinear coefficients $\alpha_{i,j,k,l}$ are proportional to $n_\mathrm{max}(t)$ (see~\autoref{eq:quartic_potential}), the Hamiltonian comprises $(2N)^4$ terms with with oscillating time-dependent prefactors. These oscillate at frequencies: $\breve{\Omega}_{i,j,k,l}^{m}=  \omega_m\pm \omega_i\pm\omega_j\pm\omega_k\pm\omega_l$. 
    
    In the optomechanical realization of the BKC, modulation frequencies are $\omega_m=\omega_i\pm\omega_j\pm\omega_k\pm\omega_l=0,\omega_i-\omega_j$ and $\omega_i+\omega_j$. Assuming resonator frequencies do not commensurate with any frequency differences or sums, the RWA retains only the so-called \textit{secular} terms that oscillate close to the frequency $\omega_j$ in the lab frame, where $\breve{\Omega}_{i,j,k,l}^{m}=0$. To filter out such contributions, we expand~\autoref{eq:3orderH} and express the result in normal order with the aid  of a symbolic algebra software~\cite{QuantumAlgebrajl_SI}: 
    %
    $H_{\text{int}}^{(3)}=\frac{1}{16}\sum_{i,j,k,l}H^{(3)}_{i,j,k,l}$. The contributions $H^{(3)}_{i,j,k,l}$ include corrections to resonator frequencies, hopping and squeezing interactions (two-body quadratic Hamiltonian contributions), and many-body interactions (three and four-body terms). Below, we categorize the secular contributions based on their degeneracy:
    \begin{enumerate}
        \item fourfold degeneracy $i=j=k=l$; single-body terms:
        \begin{align}
        H^{(3)}_{i,i,i,i} \approx \frac{\alpha_{i,i,i,i}}{16}\left[12{a}_{i}^{\dagger}{a}_{i}+6{{a}_{i}^{\dagger}}^{2}{{a}_{i}}^{2}+3\right].
        \end{align}
        Here we recover the usual form of the self-Kerr nonlinearity under the RWA.
        \item threefold degeneracy $j=k=l$, $i\neq j$; two body terms:
        \begin{align}
        H_{i,j,j,j}^{(3)}\approx\frac{3}{16}\alpha_{i,j,j,j}(1+{a}_{j}^{\dagger}{a}_{j})\left[\frac{c_{i,j}^{-}}{2}\left({a}_{i}^{\dagger}{a}_{j}+{a}_{j}^{\dagger}{a}_{i}\right)+\frac{c_{i,j}^{+}}{2}\left({a}_{i}{a}_{j}+{a}_{i}^{\dagger}{a}_{j}^{\dagger}\right)\right],
        \end{align}
        where $c_{i,j}^{\pm}$ are the modulation amplitudes for tones at $\omega_i\pm\omega_j$. Here we find corrections to linear coupling and squeezing, some of them assisted by the population of one of the modes --- a form of nonlinear gain saturation.       
        \item two-fold degeneracy $k=l$, $i=j $, $i\neq k$; three-body terms:
        \begin{align}
        H_{i,i,k,k}^{(3)}\approx\frac{\alpha_{i,i,k,k}}{16} & \left[2\left({a}_{i}^{\dagger}{a}_{i}+{a}_{k}^{\dagger}{a}_{k}\right)+2{a}_{i}^{\dagger}{a}_{i}{a}_{k}^{\dagger}{a}_{k}+1\right].
        \end{align}
        Here we find corrections to the spring effect and additional cross-Kerr interactions.
        \item simple degeneracy, $k=l$, $i\neq j $, $i\neq k$, $k\neq j$;  three-body terms:
        \begin{align}
        H_{i,j,k,k}^{(3)}\approx\frac{1}{16}\alpha_{i,j,k,k}\left(1+2{a}_{k}^{\dagger}{a}_{k}\right)\left[\frac{c_{i,j}^{-}}{2}\left({a}_{i}^{\dagger}{a}_{j}+{a}_{j}^{\dagger}{a}_{i}\right)+\frac{c_{i,j}^{+}}{2}\left({a}_{i}{a}_{j}+{a}_{i}^{\dagger}{a}_{j}^{\dagger}\right)\right].
        \end{align}
        \item All $i,j,k,l$ are different: resonator frequencies do not commensurate with any sums/substraction involving 4 frequencies, so  $\breve{\Omega}_{ijkl}^{m}\neq0$ and
        $H^{(3)}_{i,j,k,l}=0$.
    \end{enumerate}

\begin{figure*}
    \includegraphics[width=12.1cm]{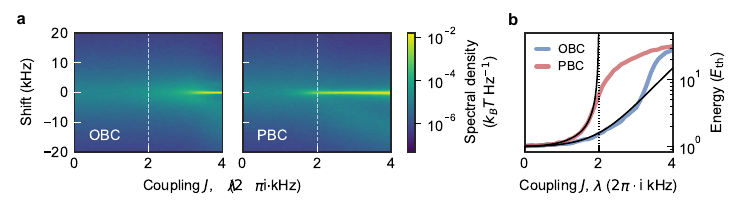}
    \caption{\textbf{Boundary-dependent instability.}
    (\textbf{a}) Thermomechanical spectra for $J = \lambda = \iu \mu$ as a function of frequency and coupling strength for periodic boundary conditions (PBC) and open boundary conditions (OBC). The coupling $\mu$ is varied over a larger range than in Fig.~2 of the main text. In addition to the dynamical instability that is predicted for PBC with standard (linear) theory, we observe a dynamic instability under OBC for large values of $\mu$ that is not predicted by linear theory. Large thermal fluctuation amplitudes then induce a nonlinear detuning that generates additional gain~\cite{McDonald2018_SI}, 
    which in turn reinforces the detuning. This sets up a positive loop for the amplification, where additionally the dynamic range of the feedback cooling system may be exceeded.
    (\textbf{b}) Total thermal energy in the chain as a function of coupling strength. For large values of $\mu$, agreement with linear theory also breaks down for OBC.
    } \label{fig:stability-spectra-extended}
\end{figure*}

\subsection{Effects of optomechanical nonlinearity on thermal populations in the presence of amplification}

In Fig. 2 of the main text, we present the detected thermal fluctuations in the BKC. The theoretical predictions of the thermal populations for both open boundary conditions (OBC) and periodic boundary conditions (PBC) is provided in section \ref{thermalsteadystates}, and compared to the experimental data in Fig. 2b for varying coupling amplitudes. As we discussed above, the optomechanical nonlinearity causes a departure of the experimentally observed signal from the linear prediction at the instability that is predicted under PBC: The optomechanical nonlinearity effectively reduces the spectral content in the detected fundamental harmonic, and the mechanical self-oscillations saturate at finite amplitude that is also determined by the nonlinearity.

In contrast to the linear theory, a dynamical instability is observed under open boundary conditions for large values of $\mu$ as well. This is shown in Fig.~\ref{fig:stability-spectra-extended}, which plots the observed fluctuations for a larger range of coupling amplitudes $|J|=|\lambda|=\mu$. In that case, thermal fluctuation amplitudes that are significantly increased due to amplification induce a nonlinear detuning (\ref{sec:optomechanical-nonlinearity}) that generates additional gain~\cite{McDonald2018_SI}, 
%
which in turn reinforces the detuning. This sets up a positive loop for the amplification, where additionally the dynamic range of the feedback cooling system may be exceeded. The fact that the departures from linear theory occurs at exactly the same recorded thermal energy under both OBC and PBC further supports the hypothesis that the onset of nonlinear effects due to the amplified motion is at the root of the observed departures.


\end{document}